\documentclass[%
 aip,
 amsmath,amssymb,
 reprint,%
]{revtex4-1}

\usepackage{graphicx}
\usepackage{dcolumn}
\usepackage{bm}
\usepackage{xcolor}
\usepackage[utf8]{inputenc}
\usepackage[T1]{fontenc}
\usepackage{mathptmx}
\usepackage{etoolbox}
\usepackage{amsmath,bm}

\usepackage{mathrsfs}
\usepackage{hyperref}
\hypersetup{colorlinks=true,allcolors=cyan}

\usepackage[left,modulo]{lineno}

\DeclareMathAlphabet{\mathcal}{OMS}{cmsy}{m}{n}
\SetMathAlphabet{\mathcal}{bold}{OMS}{cmsy}{b}{n}

\makeatletter
\def\@email#1#2{%
 \endgroup
 \patchcmd{\titleblock@produce}
  {\frontmatter@RRAPformat}
  {\frontmatter@RRAPformat{\produce@RRAP{*#1\href{mailto:#2}{#2}}}\frontmatter@RRAPformat}
  {}{}
}%
\makeatother

\begin{document}

\preprint{AIP/123-QED}

\title{Scaling particle-size segregation in wide-ranging sheared granular flows}

\author{Tianxiong Zhao \footnote{Author to who correspondence should be addressed}} 
\affiliation{School of Civil Engineering, Southeast University, Jiangning, Nanjing 211189, China}
\affiliation{Department of Earth and Environmental Science, University of Pennsylvania, Philadelphia,~PA 19104-6316,~USA.}

\author{Daisuke Noto}
\affiliation{Department of Earth and Environmental Science, University of Pennsylvania, Philadelphia,~PA 19104-6316,~USA.}

\author{Xia Li} 
\affiliation{School of Civil Engineering, Southeast University, Jiangning, Nanjing 211189, China}

\author{Tomás Trewhela}
\affiliation{Facultad de Ingeniería y Ciencias, Universidad Adolfo Ibañez, Av. Padre Hurtado 750, 2562340 Viña del Mar,~Chile}

\author{Hugo N. Ulloa$^{*}$
\footnote{Author to who correspondence should be addressed}}
\affiliation{Department of Earth and Environmental Science, University of Pennsylvania, Philadelphia,~PA 19104-6316,~USA.}
\email{ulloa@sas.upenn.edu}
\date{\today}

\begin{abstract}
Scaling relationships have been proposed to describe shear-driven size segregation based on intruder experiments and simulations~\cite{trewhela2021experimental,jing2021unified}. While these models have shown agreement with experimental and numerical results under uniform shear rate, their validity across varying shear-rate conditions remains uncertain. Here, we employ Discrete Element Method (DEM) simulations to investigate particle size segregation in sheared granular flows under wide-ranging shear-rate conditions. We find that the scaling between segregation velocity and local rheological conditions holds only within a moderate inertial number range ($0.01 < I < 0.1$), and breaks down in both quasi-static and collisional regimes. Furthermore, we show that this discrepancy leads continuum models to mispredict segregation rates in bidisperse mixtures. These findings emphasize the need for more generalized scaling laws capable of capturing segregation dynamics across a broader spectrum of shear-rate conditions and regimes.
\end{abstract}

\maketitle

\section{\label{introduction}introduction}

Particle-size segregation is a staple of many geophysical flows and industrial processes involving polydisperse grains \cite{rosato1987brazil, savage1988particle,duran1993arching, knight1993vibration,ottino2000mixing,van2015underlying,gray2018particle, umbanhowar2019review, cunez2024particle}. In natural granular flows, size segregation significantly influences sediment transport, avalanches, debris flows, and landslides \cite{felix2004relation, kleinhans2004sorting, johnson2012grain, kokelaar2014fine, de2015effects, chassagne2020discrete, kostynick2022rheology, dedieu2024sediment}. This phenomenon is also central to industrial applications such as mining, pharmaceuticals, and powder metallurgy, where the degree of mixing \cite{Trewhela_Ulloa_2024} or segregation directly impacts product quality and operational efficiency \cite{makse1997spontaneous, shinbrot2000nonequilibrium, wormsbecker2005segregation,umbanhowar2019review}. Developing predictive scaling laws for segregation dynamics is therefore crucial for optimizing and controlling granular systems in complex environments. A recent strategy to develop these scaling laws for segregation is using information from single particle (intruder) moving through a matrix of smaller or larger grains \citep{guillard2016scaling,jing2020rising,trewhela2021experimental}. While this strategy is sufficient to capture fundamental information that translates in good comparisons with experiments, there are still discrepancies that need to be addressed, including finite-concentration effects and asymmetric segregation rates\citep{bridgwater1985particle, van2015underlying,trewhela2021experimental, jing2022drag, duan2022segregation}.

Experiments and simulations consistently demonstrate that size segregation in bidisperse granular mixtures is primarily controlled by local flow conditions, notably the shear rate and overburden pressure \cite{savage1988particle, tripathi2011numerical, fan2014modelling, fry2018effect, frey2020experiments, chassagne2020discrete, trewhela2021large}.
Recently, \citet{jing2021unified} introduced a scaling relationship for intruders that differ in size or density from the surrounding particles, rooted in the competition between a buoyancy-like segregation force and a viscous drag force resisting motion. Meanwhile, Trewhela~\textit{et al.}\cite{trewhela2021experimental} took a trajectory-focused approach, directly observing the movement of intruders of varying sizes. Their work unveiled a universal scaling for segregation velocity, akin to the terminal velocity of a particle settling through a viscous fluid. These scaling relationships suggest a linear dependence between the segregation velocity and the granular flow properties. These findings underscore a fascinating analogy: particle segregation mirrors the behavior of suspended particles in viscous fluids, connecting the physics of granular flow to well-known Stokesian fluid dynamics. This analogy is natural, as granular flows can exhibit fluid-like behavior under certain external shear conditions. 

However, previous research has shown that granular materials exhibit strikingly different behaviors, transitioning between fluid- and solid-like states depending on solids volume fractions, shear and pressure conditions \cite{gdr2004dense, gaume2011quasistatic,houssais2015onset,gonzalez2023bidisperse,meng2024granular}. These transitions are governed by the inertial number, $I = \dot{\gamma}\,{d} / \sqrt{p / \rho}$, a non-dimensional parameter that encapsulates the interplay of shear rate $\dot{\gamma}$, pressure $p$, grain density $\rho$, and particle diameter ${d}$\cite{gdr2004dense, jerolmack2019viewing}. Based on $I$, granular flows are broadly classified into three regimes: the quasi-static regime, characterized by negligible grain inertia; the gaseous regime, dominated by binary collisions; and the dense flow regime, where grain inertia and enduring particle contacts coexist and influence the flow \cite{coussot1999rheophysical, gdr2004dense}. In the context of intruder segregation, recent work suggests that the driving forces on the intruder can vary with the flow regime\cite{bancroft2021drag, jing2022drag, duan2022segregation}. Such findings open questions about the broader applicability of these relationships for segregation velocity, particularly in complex, real-world granular systems.

\begin{figure*}[t!]
    \centering
    \includegraphics{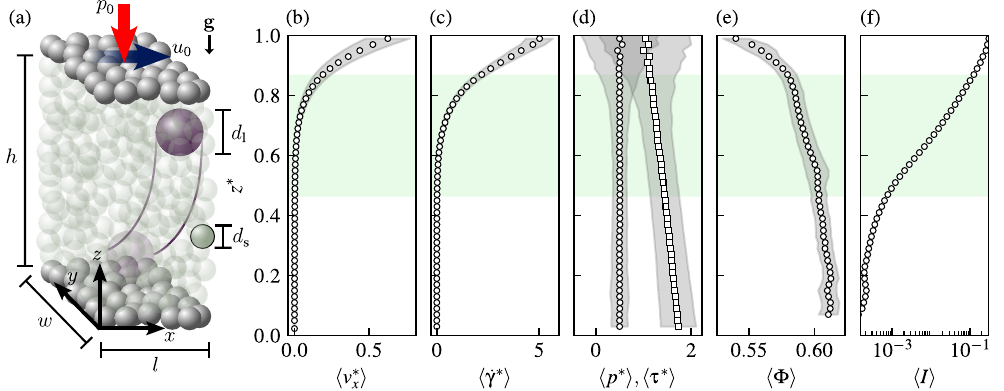}
    \caption{
    (a) Schematic representation of the DEM simulation setup in LIGGGHTS, with coordinate system ${\bm x} = (x,y,z)$. A granular bulk is sheared by a rigid upper layer---composed of particles held in fixed relative positions---moving at velocity $u_{0}$ while applying a pressure $p_{0}$ over the bulk. 
    Depth-dependent profiles of fundamental variables are obtained by ensemble averaging (denoted by $\langle \cdot \rangle$) over time, lateral directions, and cases (see \hyperref[averaging]{Appendix~\ref*{averaging}}). The gray band in each profile indicates the standard deviation from the ensemble-averaging procedure. Vertical distribution of: (b) averaged stream-wise velocity of surrounding particles $\langle {v}^*_x \rangle =\langle {v}_x \rangle/u_0$; (c) averaged shear rate $\langle \dot{\gamma}^* \rangle=\langle \dot{\gamma} \rangle/(u_0/h)$; (d) averaged pressure $\langle p^* \rangle=\langle p \rangle/p_0$, denoted by circle markers ($\bigcirc$), and averaged shear stress $\langle \tau^* \rangle=\langle \tau \rangle/p_0$, denoted by square markers ($\square$); (e) averaged solids volume fraction $\langle \Phi \rangle$; and (f) averaged inertial number $\langle I \rangle$. The highlighted green region indicates the region of the granular column where a single intruder developed segregation during the simulations. The vertical position $z$ is non-dimensionalized using the box height, $z^* = z/h$.}
    \label{fig1}
\end{figure*}

In this study, we examine the scaling of the segregation velocity for a single intruder particle subjected to depth-dependent flow conditions across a broad spectrum of inertial numbers. Specifically, we seek to answer the following question: How valid are scaling laws for particle-size segregation when applied to granular flows with non-uniform and wide-ranging inertial number conditions? Through a combination of numerical experiments and theoretical modeling, we address this question and delineate the conditions under which recent segregation relationships provide robust predictions for dynamics in complex granular flows.

This paper is structured as follows: Section~\ref{method} details the numerical methods and simulation setup used to investigate intruder segregation under non-uniform shear rate conditions. Section~\ref{background} presents the background granular flow characteristics, including depth-dependent shear rate, pressure, and inertial number distributions. Section~\ref{results} presents a comprehensive analysis of intruder trajectories and the corresponding segregation velocities for different particle size ratios and inertial numbers. Additionally, we discuss the regime-dependent nature of segregation velocity and extend the findings from the intruder case to evaluate the applicability of the introduced scaling laws in granular mixtures\cite{fry2018effect,trewhela2021experimental,jing2022drag}. Finally, Section~\ref{conclusion} summarizes the key findings and highlights potential avenues for future research.

\section{\label{method}Numerical experiments and method}

We perform a series of numerical experiments employing LIGGGHTS (LAMMPS Improved for General Granular and Granular Heat Transfer Simulations), an open-source Discrete Element Method (DEM) solver \citep{kloss2012models}.%
\textcolor{black}{\;LIGGGHTS has been benchmarked against laboratory segregation experiments, reproducing chute-flow kinematics and stratification patterns with satisfied accuracy~\citep{jiang2018influence}. To verify our simulation framework's capability to capture size-segregation dynamics, we conducted benchmark simulations reproducing established DEM results \citep{guillard2016scaling} and laboratory experiments \citep{ferdowsi2017river} (see Appendix~\ref{app:validation}). These independent validations confirm that our methodology captures essential segregation physics without ad-hoc tuning.} We use a soft-sphere model in LIGGGHTS to compute particle interactions. In this approach, particles are allowed to slightly deform upon contact, and the resulting deformations are used to calculate the contact forces. This model is particularly well suited for the quasi-static and dense granular flow conditions examined in this work \cite{zhu2007discrete}. 

These simulations aim to examine particle-size segregation in a horizontally-sheared granular medium, as illustrated in Fig.~\ref{fig1}(a). The experimental system consists of a three-dimensional granular medium subjected to gravity, defined by its dimensions $h \times l \times w$ ($h=75$ mm, $l=30$ mm, $w=10$ mm), where $h$ represents vertical height, $l$ denotes streamwise length, and $w$ denotes spanwise width. The granular sample is prepared by generating particles at randomized positions beneath the upper boundary and allowing them to settle freely under gravity. Periodic boundary conditions are applied in the planar $x$ and $y$ directions, while rigidly bonded particles are secured at the top and bottom boundaries along the $z$ direction. To drive the granular flow, the rigid top boundary moves with constant velocity $u_{0}$ in the $x$ direction, while applying an inward pressure $p_{0}$, thus transferring shear and momentum through the bulk. \textcolor{black}{This configuration produces a continuous spectrum of local inertial numbers due to depth-varying shear rate and pressure conditions that locally govern segregation dynamics. The sensitivity analysis of $u_0$ and $p_0$ conducted by \citet{trewhela2024segregation} confirms that these parameters do not qualitatively alter segregation patterns, affecting only the characteristic timescale of segregation.}

\begin{table}[ht]
\caption{\label{cases}Configuration parameters for simulation cases. Capital S and L corresponds to simulations using small and large intruders, respectively. Whereas M denotes the simulation set using granular mixtures at various average concentration $\bar{\phi}_{s}$.}
\label{cases}
\begin{ruledtabular}
\begin{tabular}{lcccc}
Case & $d_s$ ($\rm mm$) &$d_l$ ($\rm mm$) & $R$ & $\bar \phi_s$\\
\hline
S1 & $0.75$ & $1.50$ & 2.00 & $0^+$\\
S2 & $0.80$ & $1.50$ & 1.88 & $0^+$\\
S3 & $0.90$ & $1.50$ & 1.67 & $0^+$\\
S4 & $1.00$ & $1.50$ & 1.50 & $0^+$\\
S5 & $1.10$ & $1.50$ & 1.36 & $0^+$\\
L1 & $1.50$ & $2.25$ & 1.50 & $1^-$\\
L2 & $1.50$ & $3.00$ & 2.00 & $1^-$\\
L3 & $1.50$ & $4.00$ & 2.67 & $1^-$\\
L4 & $1.50$ & $5.00$ & 3.33 & $1^-$\\
L5 & $1.50$ & $6.00$ & 4.00 & $1^-$\\
M1 & $1.50$ & $3.00$ & 2.00 & $0.25$\\
M2 & $1.50$ & $3.00$ & 2.00 & $0.50$\\
M3 & $1.50$ & $3.00$ & 2.00 & $0.75$\\
\end{tabular}
\end{ruledtabular}
\end{table}

The granular bulk is composed of two types of particles of different sizes. A summary of the particle diameters and their relative proportions for each simulation case is provided in Table~\ref{cases}. The parameter $\bar{\phi}_s$ represents the average small particle concentration within the system. We first focus on the dynamics of single intruders. Here, $\bar{\phi}_s = 0^+$ (cases S1-S5) indicates a single small intruder within a bulk of larger particles, whereas $\bar{\phi}_s = 1^-$ (cases L1-L5) denotes a single large intruder within a bulk of smaller particles. Each configuration with a single intruder was replicated five times with randomized initial particle arrangements, ensuring statistical robustness. Three additional simulations (cases M1-M3), featuring intermediate values of $\bar{\phi}_s$ (0.25, 0.50, 0.75), are conducted to further examine segregation within granular mixtures. To avoid local crystallization and ensure disordered packing, the particle diameters are uniformly distributed within a 20\% deviation from the mean value \cite{pusey2009hard}.

The translational and rotational motions of individual particles are governed by Newton's second law, with the equations of motion expressed as follows \cite{kloss2012models}:
\begin{subequations}
    \begin{equation}\label{eq:linear_mom_balance}
        m_i \,\frac{d\bm{v}_i}{dt} = m_i\,\bm{g} + \sum_j \left( \bm{F}_{n,ij} + \bm{F}_{t,ij} \right),
    \end{equation}
    \begin{equation}\label{eq:angular_mom_balance}
        I_i \,\frac{d\boldsymbol{\omega}_i}{dt} = \sum_j \bm{r}_{ij} \times \bm{F}_{t,ij}, 
    \end{equation}
\end{subequations}
where $m_i$ and $I_i$ are the mass and moment of inertia of particle $i$, $\bm{v}_i$ and $\boldsymbol{\omega}_i$ are its translational and rotational velocities, $\bm{g}$ is the gravitational acceleration, and $\bm{r}_{ij}$ is the vector from the particle center to the contact point. $\bm{F}_{n,ij}$ and $\bm{F}_{t,ij}$ represent the normal and tangential contact forces between the particles $i$ and $j$. These contact forces are calculated using an elastic spring-dashpot model, and they are expressed as:
\begin{subequations}
\begin{equation}\label{eq:normal_force}
\bm{F}_{n,ij} = k_n \delta_{n,ij} - \gamma_{n,ij} \dot{\delta}_{n,ij}, 
\end{equation}
\begin{equation}\label{eq:tangential_force}
\bm{F}_{t,ij} = k_t \delta_{t,ij} - \gamma_{t,ij} \dot{\delta}_{t,ij},
\end{equation}    
\end{subequations}
where $k_n$ and $k_t$ are the normal and tangential stiffness, $\delta_{n,ij}$ and $\delta_{t,ij}$ are the normal overlap and tangential displacement, $\gamma_{n,ij}$ and $\gamma_{t,ij}$ are the damping coefficients, $\dot{\delta}_{n,ij}$ and $\dot{\delta}_{t,ij}$ are the normal and tangential deformation rate of particles $i$ and $j$, respectively. The tangential force is capped by a Coulomb friction criterion, $|\bm{F}_{t,ij}| \leq \mu_p |\bm{F}_{n,ij}|$, where $\mu_p$ is the particle friction coefficient.%
\textcolor{black}{\;The normal and tangential stiffness values ($k_n$ and $k_t$) are selected consistent with the previous segregation study by Guillard~et~al. \cite{guillard2016scaling}.}%
This approach captures the elastic deformation, energy dissipation, and frictional interactions between particles, enabling accurate modeling of dense granular flows under shear. The values for the numerical parameters used in our simulations, including stiffness, damping coefficients, and friction, are summarized in Table~\ref{parameters}. \textcolor{black}{Systematic DEM studies report that moderate variations of stiffness or density shift segregation indices by at most a few percent~\citep{dziugys2007influence}, whereas varying the inter-particle friction coefficient rescales the rate of segregation without altering the qualitative trend~\citep{li2022dem}. Hence the trends reported here are robust to the precise parameter set chosen.}

\begin{table}[h]
\caption{\label{parameters}Numerical constants and parameters in DEM simulations.}
\label{parameters}
\begin{ruledtabular}
\begin{tabular}{lcc}
Parameter & Symbol & Value \\
\hline
Normal stiffness & $k_n$ & $8.1 \times 10^5$ N/m \\
Tangential stiffness & $k_t$ & $8.6 \times 10^5$ N/m \\
Friction coefficient & $\mu_p$ & 0.5 \\
Particle density & $\rho$ & 2500 kg/m$^3$ \\
Top boundary velocity & $u_0$ & 2.5~m/s \\
Top boundary pressure & $p_0$& 1600~Pa \\
Tangential damping & $\gamma_t$ & 12 kg/s \\
Normal damping & $\gamma_n$ & 12 kg/s \\
Domain height & $h$ & $75$ mm \\
Domain length & $l$ & $30$ mm \\
Domain width & $w$ & $10$ mm \\
Timestep & $\Delta t$ & $10^{-6}$ s \\
\end{tabular}
\end{ruledtabular}
\end{table}

\section{\label{background}Background granular flow}
The background granular flow characteristics are obtained by averaging fundamental quantities over time and laterally across the domain, which we denote by $\langle\xi\rangle$, with $\xi(t,{\bm x})$ representing an arbitrary field. Notice that $\langle\xi\rangle$ is a depth-dependent quantity, and for generality, they are expressed in non-dimensional form, represented as $\xi^{*}$. Figure~\ref{fig1}(b) presents the normalized streamwise velocity profile, $\langle v_x^*\rangle= \langle v_x\rangle / u_0$, where $u_0$ is the velocity of the top boundary [Table~\ref{parameters}]. Considering the background flow structure, the shear rate $\dot{\gamma}$ is determined by the vertical gradient of the streamwise velocity, $\dot{\gamma}=\partial v_x/\partial z$, which exhibits significant variation with granular column depth. Its vertical distribution in non-dimensional form  $\langle\dot\gamma^*\rangle= \langle\dot{\gamma}\rangle/(u_{0}/h)$ is shown in Fig.~\ref{fig1}(c). The general trend of the vertical shear rate can be represented by an exponential function $\dot{\gamma}(z) \propto \gamma_0 \,\exp({z/\lambda})$, where $z$ is the vertical coordinate representing height and $\lambda$ is the characteristic $e$-folding length scale. This vertical exponential distribution of the shear rate is consistent with the findings and theory documented in the literature\cite{mueth2003measurements, may2010shear, artoni2018shear,chassagne2020discrete}.

Figure~\ref{fig1}(d) shows the vertical distribution of the computed non-dimensional pressure $\langle p^* \rangle= \langle p \rangle/p_{0}$ and the shear stress $\langle \tau^* \rangle=\langle \tau \rangle/p_{0}$ based on the sum of the contact forces between the particles above and below a given height. Specifically, the contact forces were decomposed into vertical and streamwise components and divided by the cross-sectional area $A=w \times l$ [Fig.~\ref{fig1}(a)] to estimate the pressure and the shear stress as a function of the non-dimensional height $z^*$. Additionally, both quantities were normalized by the pressure exerted at the top boundary $p_{0}$. As expected and in agreement with previous studies\cite{guillard2016scaling}, the pressure $\langle p^{*} \rangle$ (circles) increases almost linearly from top to bottom due to the cumulative weight of the granular column. Whereas the shear stress $\langle \tau^{*} \rangle$ (squares) remains nearly uniform in the vertical direction. 

Due to the varying flow conditions within the granular system, we observe a depth-dependent solids volume fraction $\langle \Phi \rangle$ [Fig.~\ref{fig1}(e)]. The variable $\langle \Phi \rangle$ distribution reveals a denser particle packing near the bottom boundary, where the reduced velocities and shear rates minimize particle movement and rearrangement. In contrast, lower packing fractions are observed near the top boundary, associated with higher dilation and shear rates, which in turn favor particle movement. The solids volume fraction can be described using the equation $\Phi(z) = \Phi_b - \Phi_a (1 - {\rm exp}({z/\theta}))$, where \(\Phi_b\) is the volume fraction at the bottom of the granular column ($z=0$), $\Phi_a$ quantifies the deviation of \(\Phi\) from \(\Phi_b\) as \(z\) increases, and \(\theta\) is the characteristic length scale over which \(\Phi(z)\) transitions from its bottom value. 

Inheriting from the variable shear rate $\langle \dot{\gamma} \rangle$ and pressure $\langle p \rangle$ distributions, a local $z$-dependent inertial number $\langle I \rangle= \langle\dot{\gamma}\rangle \bar{d}/\sqrt{\langle p \rangle/\rho}$ is obtained. Here, $\rho$ represents particle density, and $\bar{d} = \phi_s d_s+\phi_l d_l$ is the concentration averaged particle diameter, where $\phi_s$ and $\phi_l$ denote the volume fractions of small and large particles per unit granular volume respectively. This definition of $\bar{d}$ is not only intuitive to deal with bidisperse mixtures~\cite{rognon2007dense,tripathi2011numerical}, it also naturally introduces the asymmetric coefficient for segregation $\chi$~\cite{van2015underlying}, i.e. $\bar{d}=(1-\chi\phi_{s})Rd_{s}$~\cite{trewhela2021conveyor,trewhela2024segregation}. As a result, the inertial number $I$ decreases from higher values near the top, $I \approx 0.25$, to lower values near the almost motionless bottom boundary, $I \approx 4 \times 10^{-4}$. These features underscore the interplay between shear and gravity in shaping the granular bulk, resulting in nonlinear dynamics and depth-dependent flow properties, in agreement with previous studies on sheared granular systems\cite{may2010shear,chassagne2020discrete, mueth2003measurements, artoni2018shear, pouliquen2006flow}. The lower region of the granular column exhibits a more stable stationary state, behaving like a sheared solid with minimal particle motion. In contrast, as the shear rate increases toward the top, the granular material transitions into a more fluid-like, dynamic regime characterized by frequent particle collisions and rearrangements. This pronounced vertical gradient in flow behavior profoundly influences the dynamics of both intruder particles and their surrounding granular matrix, emphasizing the importance of examining intruder behavior from a local perspective.

\section{\label{results}results and discussion}

\subsection{Intruder trajectory}

\begin{figure}[h]
    \centering
    \includegraphics{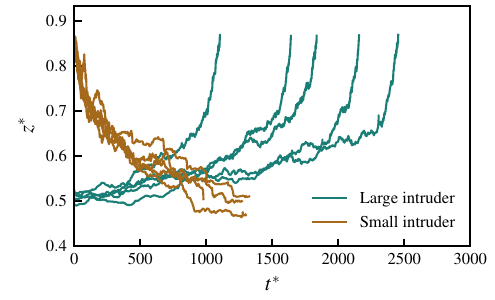}
    \caption{Intruders trajectories with value $R=2$ as a function of non-dimensional time $t^*=t \, (u_0/h)$. The diagram distinguishes between the large rising intruder, represented in green, and the small sinking intruder, depicted in brown.}
    \label{fig2}
\end{figure}

Considering the low shear rates observed in the deeper region [Fig.~\ref{fig1}(b)], particle-size segregation is either negligible or occurs at an extremely slow rate. To capture the segregation process within a shorter simulation time and reduce costs, our analysis of the segregation dynamics of a single intruder concentrates on regions where the non-dimensional shear rate $\langle \dot\gamma^*\rangle$ falls within the range of $10^{-3}$ to $10^{-1}$, which corresponds to height $z^{*} = z/h$ values between 0.47 and 0.87, as highlighted by the shaded green areas in Fig.~\ref{fig1}. The intruder is analyzed by tracking its vertical trajectory over time. Figure~\ref{fig2} presents the temporal evolution of vertical positions for small and large intruders in separate experiments using the same size ratio, $R=2$. The small intruder is initially positioned near the top of the bed ($z^*\approx 0.9$), and the large intruder starts at $z^*\approx 0.45$. The results reveal that the small intruder percolates downward more rapidly than the large intruder rises upward, demonstrating an asymmetric behavior consistent with previous experimental findings \cite{van2015underlying,trewhela2021experimental}. This faster and more erratic motion is likely due to the small intruder's ability to fall through the larger gaps created by the rearrangement of larger surrounding particles. 

To further quantify the segregation dynamics, we compute the vertical velocity of the intruder particle during its movement through the granular medium, $w_\text{seg}$. Calculating the velocity of the intruder particle directly from its vertical position, $z_p(t)$, by calculating the derivative $\mathrm{d}z_{p}/{\mathrm{d}t}$, would yield highly noisy results due to the inherent scatter in the time-integrated trajectories, as illustrated in Fig.~\ref{fig2}. To mitigate this issue and obtain a meaningful segregation velocity scale, we adopt an alternative approach by analyzing the time required for the intruder to travel a specific vertical distance $d_\text{cp}$. We denote the heights characterizing the start and end of these vertical sections as checkpoints.

\begin{figure}[h]
    \centering
    \includegraphics{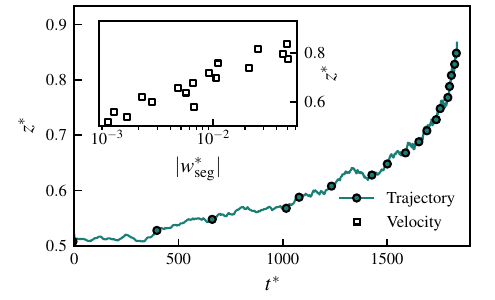}
    \caption{Trajectory, checkpoint, and intruder velocity for a representative case ($R = 2$). 
    The main plot shows the intruder position over $t^*$, with the curves representing the intruder's trajectory and the markers indicating checkpoints used to measure the time the intruder takes to travel $d_\text{cp}$. The inset shows the corresponding $|w^*_{\text{seg}}|$ as a function $z^*$.}
    \label{fig3}
\end{figure}

The choice of the vertical distance $d_\text{cp}$ is critical and should reflect the underlying mechanism driving the segregation process. \citet{van2015underlying} found that the transition length scale between the diffusive and segregation (superdiffusive) regimes is approximately equal to the diameter of the surrounding large particles for small intruders. For large intruders, this length scale is smaller as the intruders tend to be more stable and exhibit fewer fluctuations~\cite{trewhela2021experimental}. Here, we select $d_\text{cp} = \bar{d}$ as the distance between adjacent checkpoints, as illustrated in Fig.~\ref{fig3} where the trajectory of a large intruder of size ratio $R$ is plotted with its corresponding checkpoints. This scale is selected to be sufficiently large to be in the superdiffusive regime, minimizing the impact of transient fluctuations in the granular medium. The particle velocity is then estimated as $w_{\text{seg}} \approx d_\text{cp} / \Delta t$, where $\Delta t$ is the time interval required for the intruder to traverse the distance $d_\text{cp}$.

The inset of Fig.~\ref{fig3} illustrates the magnitude of the non-dimensional intruder velocity $|w_{\text{seg}}^*| = |{w_{\text{seg}}}|/{\sqrt{g \, \bar{d}}}$ as a function of the position $z^*$, following the non-dimensionalization approach used in \citet{chassagne2020discrete}. Both the trajectory and the segregation velocity depicted in Fig.~\ref{fig3} point towards a strong influence of the local shear-rate condition in the intruder's movement within the granular column. In the lower shear rate regime, corresponding to the deeper regions of the system, the intruder moves slowly, with longer times between checkpoints and a lower segregation velocity. Conversely, the segregation velocity is significantly larger when the intruder is in upper regions, where shear rate is greater.

\subsection{Segregation velocity}

To obtain a more rigorous, quantitative description of the intruder’s motion, we adopt a mixture-theory-based continuum framework and  treat the intruder as one distinct component within the granular medium \cite{gray2018particle}. The spatiotemporal distribution of large and small particles in a bidisperse mixture can be described by incorporating segregation fluxes and diffusive remixing: 
\begin{subequations}\label{eq:continuum_model}
\begin{equation}
    \frac{\partial \phi^l}{\partial t} + \nabla \cdot (\phi^l \mathbf{u}) - \nabla \cdot \left( f_{sl} \phi^l \phi^s \frac{\mathbf{g}}{|\mathbf{g}|} \right) = \nabla \cdot (D_{sl} \nabla \phi^l),
    \label{eq:phi_l}
\end{equation}
\begin{equation}
    \frac{\partial \phi^s}{\partial t} + \nabla \cdot (\phi^s \mathbf{u}) + \nabla \cdot \left( f_{sl} \phi^s \phi^l \frac{\mathbf{g}}{|\mathbf{g}|} \right) = \nabla \cdot (D_{sl} \nabla \phi^s).
    \label{eq:phi_s}
\end{equation}
\end{subequations}
Here, $\phi_l$ and $\phi_s$ denote the proportions of large and small particles within the granular mixture, defined as: $\phi_s = \Phi_s / \Phi$ and $\phi_l = \Phi_l / \Phi$, where $\Phi_s$ and $\Phi_l$ are the volume fractions of small and large particles, respectively.
 $\mathbf{u}$ is the bulk granular velocity field, $\bm{g}$ is the gravitational acceleration vector, and $D_{sl}$ is the diffusion coefficient.  $f_{sl}$ is the segregation velocity magnitude. Under cyclic shear conditions, Trewhela \textit{et al.}\cite{trewhela2021experimental} demonstrated that $f_{sl}$ scales inversely with local pressure and linearly with the shear rate:
\begin{equation}\label{eq:fsl}
    f_{sl} = \mathcal{F}(R) \, \left(\frac{\dot{\gamma}}{p}\right)\, \rho \, g \, \bar{d}^2,
\end{equation}
where $\mathcal{F}(R)$ is a function reflecting the role of the particles' size ratio $R$ on the segregation dynamics. In conditions where the contribution of diffusion is negligibly small, the vertical velocity $w_\text{seg}$ of species $i$ (large or small particles) is given by:
\begin{equation}
    w_{\text{seg},i} = \pm f_{sl}(1 - \phi_i),
\end{equation}\label{eq:wseg}
where the $+$ sign applies to large particles (upward velocity) and the $-$ sign to small particles (downward velocity). Here, $(1 - \phi_i)$ represents the volume fraction of the opposing particle species. For instance, in the case of a single large intruder, where $\phi_l= 0^+$, segregation velocity reduces to $w_{\text{seg}} = f_{sl}$. In this study, we compute a local normal viscosity $\eta_\text{nom} = p/\dot{\gamma}$ as an effective descriptor of the granular rheology.

\begin{figure}[htbp]
    \centering
    \includegraphics{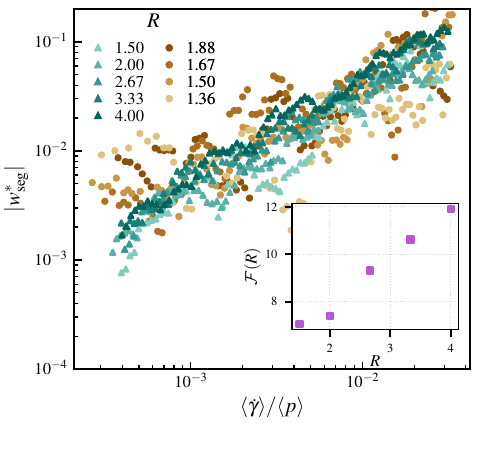}
    \caption{\textcolor{black}{Non-dimensional intruder velocity magnitude$|w_{\text{seg}}^*|$ as a function of $\eta_\text{nom}^{-1}=\langle \dot{\gamma} \rangle/\langle p \rangle$. Green colors represent large rising intruder cases, while brown colors represent small sinking ones. Deeper colors correspond to cases with higher size ratios $R=d_l/d_s$. The inset shows the size ratio function $\mathcal{F}(R)$ with error bars representing standard deviation.}}
    \label{fig4}
\end{figure}

\begin{figure*}[htbp]
    \centering
    \includegraphics{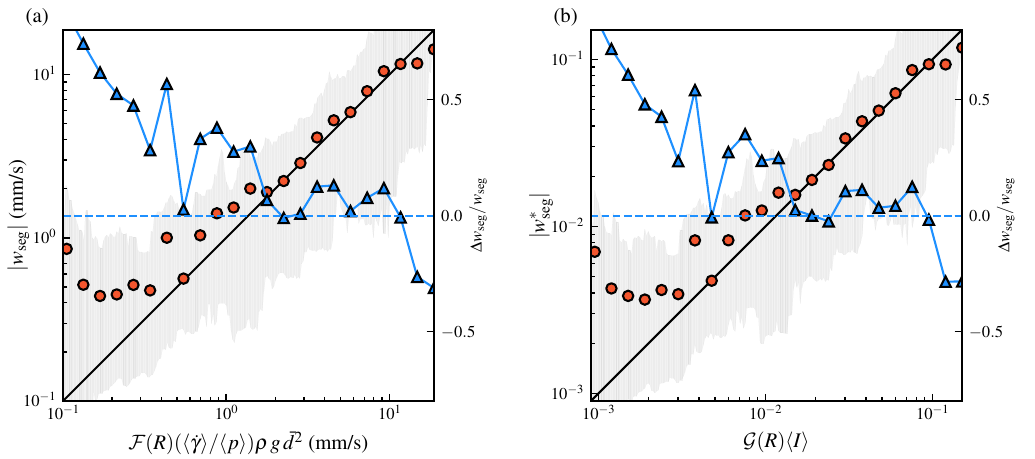}
\caption{The magnitude of the intruder velocity $| w_{\text{seg}}|$ plotted against the scaled variables in sheared granular flows. (a) The magnitude of the intruder velocity $| w_{\text{seg}}|$ versus $\mathcal{F}(R) \, (\langle\dot{\gamma}\rangle/\langle p \rangle) \rho g \bar{d}^2$, with the best-fit function $\mathcal{F}(R) = A_1 (R-1)^{B_1}$ yielding $A_1 = 7.75 \pm 0.28$ and $B_1 = 0.19 \pm 0.05$. (b) The magnitude of the non-dimensional intruder velocity, $|w_{\text{seg}}^*|$, plotted against the scaled inertial number $\mathcal{G}(R) \, \langle I \rangle$, with $\mathcal{G}(R) = A_2 (R-1)^{B_2}$ and best-fit parameters $A_2 = 1.08 \pm 0.03$ and $B_2 = 0.18 \pm 0.04$. The shaded gray bands indicate the 10th to 90th percentile ranges of the measured velocities. The orange circles represent binned mean values, and blue triangles show the relative error $\Delta w_{\text{seg}}/w_{\text{seg}}$, which quantifies the deviation between the measured segregation velocity and the corresponding prediction from the corresponding scaling model.}
\label{fig5}
\end{figure*}

Figure~\ref{fig4} displays scattered points representing the intruder velocity magnitude $|w_{\text{seg}}^*|$ at different heights $z^*$ as a function of the local $\eta_\text{nom}^{-1}$ calculation. The points are color-coded to indicate data from different size ratios $R$ and cases of large or small intruders. The results show a general trend with a slope close to unity on the logarithmic scale, indicating that the relationship $f_{sl} \propto (\dot{\gamma} / {p})$ holds fairly well. However, deviations occur in regions with low shear rate, where small intruders exhibit faster segregation velocities than expected by the linear trend. This asymmetry reveals that small intruders move more efficiently within a matrix of larger particles than large intruders in a matrix of smaller particles. While small intruders just rely on gaps to fall through the granular matrix, large particles rely on more complex mechanisms that are still poorly understood \citep{van2018segregation,trewhela2021large,dedieu2024sediment}. Furthermore, the segregation velocity of small intruders is more scattered than that of large intruders, likely due to the greater influence of local random packing configurations on their ability to percolate through gaps between larger particles. Here, we limited the size ratio of small intruders to values below $R = 2$, as larger ratios tend to produce increasingly erratic dynamics and reduced reproducibility due to the emergence of spontaneous percolation mechanisms at large $R$. Despite the fact that both packing and segregation mechanisms are matters of debate and active research, single intruder trajectories and velocities from our DEM results provide sufficient information to feed state-of-the-art models.

The velocities corresponding to different size ratios shown in Fig.~\ref{fig4} exhibit a systematic variation with $R$, indicating the influence of the size difference on the segregation process. Within the range of size ratios studied, a larger $R$ leads to faster segregation velocities. This influence of $R$ on segregation expressed as a function $\mathcal{F}(R)$ should ensure that the segregation velocity vanishes when $R = 1$. Based on this constraint and the simulation results, we propose a minimal functional form for $\mathcal{F}(R)$ that collapses the segregation velocity data across different size ratios:
\begin{equation}\label{eq:FR}
    \mathcal{F}(R) = A_1 (R - 1)^{B_1}.
\end{equation}

\begin{figure*}[ht]
    \centering
    \includegraphics{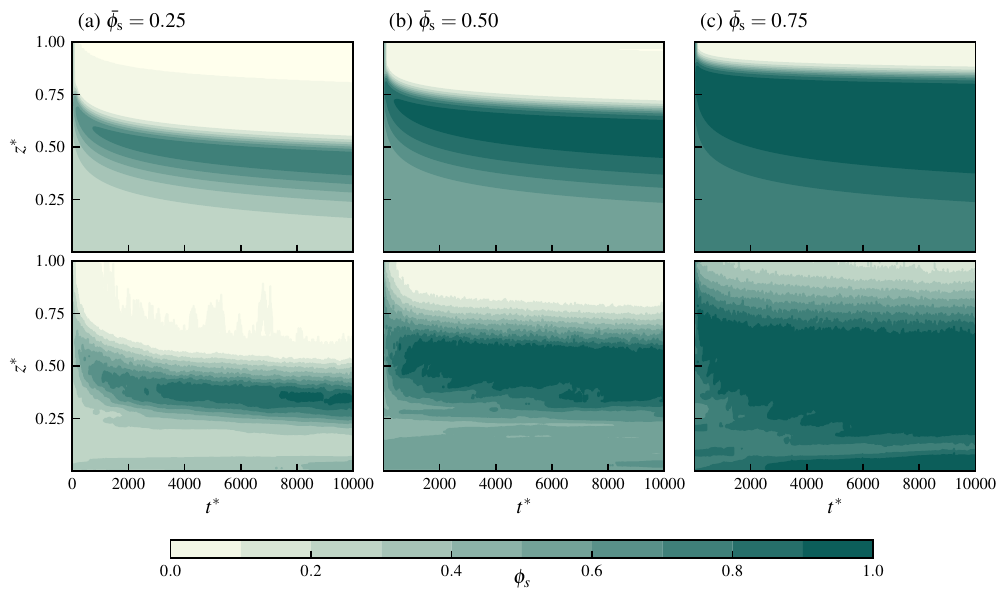}
    \caption{Evolution of the vertical distribution of small particles concentration $\phi_s$ over time in a bidisperse systems considering three overall (averaged) small particle volume fractions $\bar{\phi_s}$. The upper row presents results from the continuum model (\ref{eq:small_particles}) incorporating the segregation velocity rate scaling (Eq.~\ref{eq:fsl}); whereas the second row presents DEM simulation results (see appendix~\ref{DEM mixture analysis}). Each column corresponds to an overall small particle volume fraction: $\bar{\phi_s} = 0.25$, 0.50, and 0.75. The horizontal axis represents the non-dimensional time ($t^*=t\,U_{0}/h$) and the vertical axis represents the non-dimensional height ($z^{*}=z/h$). The color scale represents the volume fraction of small particles $\phi_s$, with lighter colors indicating lower concentrations and darker colors indicating higher concentrations.} 
    \label{fig6}
\end{figure*}

To determine the best fit values for the coefficient $A_1$ and the power exponent $B_1$, we grouped the $|w_{\text{seg}}|$ for the corresponding values of $R$, and performed a least squares optimization. The process involves minimizing the squared differences between the measured $|w_{\text{seg}}|$ and the values predicted by equation \ref{eq:fsl} using the local $\langle \dot{\gamma} \rangle$ and $\langle p \rangle$ distributions (see \hyperref[averaging]{Appendix~\ref*{FRfitting}}). Figure~\ref{fig5}(a) displays $|w_{\rm seg}|$ against the scaling for the segregation velocity magnitude $f_{sl}$ in (\ref{eq:fsl}), The shaded gray band represents the 10- to 90-percentile range of the measured particle segregation velocities, whereas the red circles denote the mean values. The blue triangles indicate the relative error $\Delta w_{\rm seg}/w_{\rm seg}$ quantifying the deviation between the DEM simulation data and the scaling prediction. It is apparent that $|w_{\rm seg}|$ follows a linear dependence with the scaling term $\mathcal{F}(R) \, \left({\dot{\gamma}}/{p}\right)\,\rho g \bar{d}^2$ over a specific range of the scaling variable. Beyond this range, the scaling predictions diverge from the DEM simulation data: The segregation velocity is underestimated for low $\langle \dot{\gamma} \rangle/\langle p \rangle$ and overestimated for high $\langle \dot{\gamma} \rangle/\langle p \rangle$. The segregation scaling velocity proposed by \citet{jing2022drag}, which has a similar form but characterizes the effective viscosity through the local shear stress $\tau$ rather than the pressure $p$, is also tested and yields a similar trend (see appendix~\ref{scaling_Jing}). This deviation confirms the reduced influence of the shear rate condition on $w_{\text{seg}}$ in the low shear rate regime, where other factors, such as local packing configurations, particle material properties and diffusion, may play a more dominant role. Discrepancies in the high shear rate region, meanwhile, suggest that inertial effects or collisional dissipation suppress segregation efficiency, overriding the shear-rate-driven dynamics assumed by the scaling frameworks. These results suggest that the effectiveness of linear shear rate dependent scaling models may significantly decline outside intermediate or transitional flow regimes.

Previous research has demonstrated that many granular flow characteristics --- such as velocity profiles, flowing layer depth, and volume fraction --- depend strongly on the inertial number \cite{gdr2004dense, jop2006constitutive, chassagne2020discrete}. Motivated by the dependence of granular flow behavior on the inertial number, we also scale the non-dimensional intruder velocity using $I$: 
\begin{equation}\label{eq:w_seg_I}
\frac{|w_{\rm seg}|}{\sqrt{g\,\bar{d}}} = \mathcal{G}(R) \, I,
\end{equation}
which resembles the scaling suggested in previous DEM studies~\cite{chassagne2020discrete, bancroft2021drag, fry2018effect}. Here, $\mathcal{G}(R)$ is a function that accounts for the effect of the size ratio $R$. In Fig.~\ref{fig5}(b), the non-dimensional intruder velocity, $|w_{\rm seg}|/\sqrt{g\,\bar{d}}$, is plotted against $\langle I \rangle$. The trend in Fig.~\ref{fig5}(b) is similar to that in panel (a): a clear linear regime is observed over a specific range of $\langle I \rangle$, while deviations occur outside this range. These observations imply that the scaling relation for the segregation velocity of intruder particles is most applicable when $\langle I \rangle$ lies between approximately 0.01 and 0.1. Outside this regime, deviations become apparent, suggesting that additional factors, such as enhanced collisional dissipation and increased particle diffusion, exert a stronger influence on particle motion.

\subsection{Segregation in mixture cases}

We now extend our investigation to mixture cases to explore how deviations and predictions translate to bidisperse granular mixtures. These additional simulations are built on the results from the previous subsection, by introducing the fitted parameters into continuum theoretical segregation models for bidisperse mixtures~\cite{gray2018particle}.

To simplify the governing equations, we assume that the particle size distribution is spatially uniform along the $x$- and $y$-directions. The segregation equation for small particles simplifies to a reduced form:
    \begin{equation}\label{eq:small_particles}
     \frac{\partial \phi_s}{\partial t} -\frac{\partial}{\partial z} \left(f_{sl} \phi_l \phi_s\right) = \frac{\partial}{\partial z} \left(D_{sl} \frac{\partial\phi_s}{\partial z}\right),
    \end{equation}
where $f_{sl}$ is the segregation velocity magnitude following Eqs.~\eqref{eq:fsl} and~\eqref{eq:FR}, whereas $D_{sl}$ is the diffusivity calculated as $D_{sl} = \mathcal{A}\,\langle\dot{\gamma}\rangle\,\bar{d}^2$, with $\mathcal{A}=0.108$ \cite{utter2004self}, previously used as a diffusivity proxy for continuum simulations and numerical solutions~\cite{trewhela2024segregation,maguire2024particle}. The shear rate profile of the background granular flow (see section~\ref{background}) is modeled by the function
$\langle\dot{\gamma}\rangle = (u_0/\lambda) \exp(z/\lambda)$, where the fitted parameters $u_0 = (9.49 \pm 5.00) \times 10^{-6}~\mathrm{m/s}$ and $\lambda = (6.09 \pm 0.27)\times 10^{-3}~\mathrm{m}$ were obtained by fitting the least squares of the velocity profiles ($R^2 = 0.976$).

To evaluate the performance of the continuum model (CM), DEM numerical experiments for bidisperse mixtures were carried out and used for comparison. In this third set of simulations, we used the same configuration as for the intruder simulations, i.e. we maintained identical boundary conditions and particle properties. However, we kept the particle size ratio $R=2$ fixed so that we focus on and compare three different cases of the overall small particle concentration $\bar{\phi_s}$: 25\%, 50\%, and 75\%. Concurrently, we solved numerically the nonlinear partial differential equations in \eqref{eq:continuum_model} using the method of lines\cite{trewhela2024segregation}.

Figure~\ref{fig6} illustrates the evolution of the local volume fraction of small particles $\phi_s$, across the granular column ($z^*$) over time ($t^*$). Please note that large particle concentrations are not plotted for convenience and conciseness, since $\phi_{l}=1-\phi_{s}$. Although some discrepancies are observed between the CM and DEM simulations, the model incorporating the segregation velocity scaling (Eq.~\ref{eq:fsl}) captures well the general trends in the evolution of $\phi_s$ across all three cases. Notable differences arise in cases with either a high ($\bar{\phi_s} = 75\%$) or low ($\bar{\phi_s} = 25\%$) fraction of small particles. In these cases, the DEM simulations indicate that some small particles have already reached the bottom region, a behavior not predicted by the CM. Moreover, the DEM results show more diffusivity than the CM. This diffusion is observed in the blurred transitions between high and low small particles concentrations. These blunt transitions are especially apparent in the case of $\bar{\phi}_{s}=0.75$ for the average small particle concentration (Fig.~\ref{fig6}c). This discrepancy is consistent with the intruder segregation results, where the linear velocity formulation tends to underestimate segregation in low shear rate regions, particularly when small particles percolate through a matrix of larger ones.

\begin{figure}[htbp]
    \centering
    \includegraphics{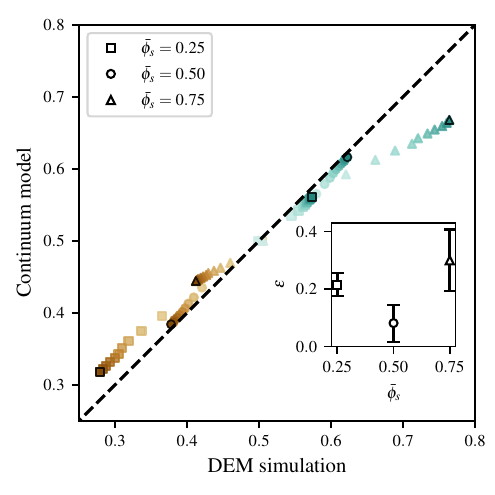}\label{fig7}
    \caption{Center of mass (COM) of small and large particles in DEM simulations and continuum model for bidisperse granular mixtures. Marker shapes denote initial solid volume fractions ($\bar{\phi}_s = 25\%$\,$\square$, $\bar{\phi}_s = 50\%$\,$\bigcirc$, $\bar{\phi}_s = 75\%$\,$\triangle$). Marker colors reflect particle type: small particles (orange) and large particles (green). Color intensity indicates normalized temporal progression (lighter: initial states, darker: final equilibrium). Dashed line indicates theoretical agreement ($y = x$). The inset shows the relative error in center of mass $\varepsilon$ as a function of $\bar{\phi}_s$. Error bars represent the standard deviation over the simulation time span.}
    \label{fig7}
\end{figure}

We quantify the difference between the DEM experiments and the CM results by tracking the vertical coordinate of the center of mass (COM) of each particle fraction (small and large) as an indicator of the segregation progression (Fig.~\ref{fig7}). The domain is vertically divided into discrete bins of $3 \, \text{mm}$ size, equal to the diameter of the large particles in the mixture. The COM for each particle species $r$ (small or large) at time $t$ is calculated as:
\begin{equation}
    \text{COM}_r(t) = \frac{\sum_{k} z_k \, \phi_{r,k}(t)}{\sum_{k} \phi_{r,k}(t)},
    \label{eq:com}
\end{equation}
where $z_k$ is the vertical position of the center of bin $k$, and $\phi_{r,k}$ is the volume fraction of particle species $r$ in bin $k$. Markers in Fig.~\ref{fig7} are colored by particle types, brown and green for small and large particles, respectively. The intensity of the color reflects the simulation time, transitioning from light (initial stages) to dark (final stages). The inset in Fig.~\ref{fig7} shows the error $\varepsilon(t)$ of the CM relative to the DEM simulations, which is defined as:
\begin{equation}
    \varepsilon(t) = \frac{\Delta z_{\mathrm{DEM}}(t) - \Delta z_{\mathrm{CM}}(t)}{\Delta z_{\mathrm{DEM}}(t)}.
    \label{eq:error}
\end{equation}
where $\Delta z_{\rm DEM}(t) = \text{COM}_{\rm DEM}(t) - \text{COM}_{\rm DEM}(0)$ and $\Delta z_{\rm CM}(t) = \text{COM}_{\rm CM}(t) - \text{COM}_{\rm CM}(0)$ are the displacements of the COM for the DEM and CM results relative to their initial position, respectively. In general, the results demonstrate that the CM underestimates the rate of segregation. The intermediate mixture case ($\bar{\phi}_s = 0.50$) exhibits the lowest deviation ($\varepsilon = 0.081 \pm 0.064$), while larger discrepancies occur in the $\bar{\phi}_s = 0.25$ ($\varepsilon = 0.215 \pm 0.040$) and $\bar{\phi}_s = 0.75$ ($\varepsilon = 0.300 \pm 0.107$) cases. This tendency of the CM to underestimate the extent of segregation is observed in all three cases, and consistent with observations from the intruder cases. The comparison highlights that, while the CM captures well the direction and rate of segregation observed in the DEM simulations, its reliance on the linear velocity scaling can lead to significant differences in mixture cases.

\begin{figure*}[t!]
    \centering
    \includegraphics{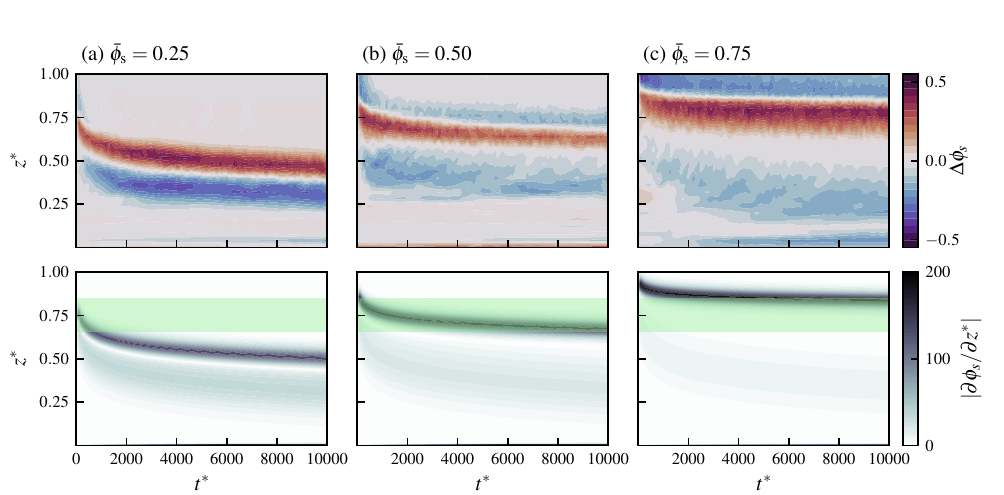}
    \caption{Comparison of segregation dynamics in bidisperse granular mixtures between DEM simulations and CM predictions for the three averaged small particle concentration cases: $\bar{\phi}_\mathrm{s} = 0.25$, $0.50$, and $0.75$. Top row shows the difference in small particle concentration $\Delta\phi_s = \phi_{s,\text{model}} - \phi_{s.\text{DEM}}$. Bottom row presents evolution of the composition interface quantified by the absolute vertical gradient of small particle volume fraction, $|\partial\phi_s/\partial z^*|$, for three initial mixture configurations. Color intensity shows gradient magnitude. The light green band highlights the inertial number range $0.01 < \langle I\rangle < 0.1$.}
\label{fig:interface_gradients}
\end{figure*}

As large particles rise and small particles sink in a bidisperse mixture, a distinct composition interface forms between these segregating species. This interface region, characterized by a steep gradient in small-particle concentration, delimits the boundary between the small-particle-rich zone and the large-particle-rich zone. It indicates the central zone of interparticle flux, where both segregation and diffusion act to redistribute particles. Figure~\ref{fig:interface_gradients} shows contour plots of the vertical gradient of the small particle volume fraction, $\left|\partial \phi_s/\partial z\right|$, over time for three different initial mixture compositions ($\bar{{\phi}}_s = 0.25$, 0.50, and 0.75). The dark band indicates the location of the composition interface that separates the small- and large-particle-dominated regions. A light green horizontal band highlights the regime where the local inertial number $I$ is between 0.01 and 0.1, for which we previously showed that the scaling law for the segregation velocity is most accurate.

The spatiotemporal evolution of the segregation interface (Fig.~\ref{fig:interface_gradients}) critically determines the validity of the predictions made by the continuum model. When the interface resides within the moderate inertial number regime ($0.01 <\langle I \rangle< 0.1$, highlighted in light green), the continuum model achieves strong agreement with the DEM results. This agreement is particularly evident in the case $\bar{\phi}_s = 0.50$, where the interface remains within this optimal vertical band for most of the segregation process. In contrast, systems with extreme compositions ($\bar{\phi}_s = 0.25$ and $0.75$) exhibit interfaces that spend significant time outside the validated inertial number range. For $\bar{\phi}_s = 0.75$, the interface rises into the high-$\langle I \rangle$ collisional regime, while for $\bar{\phi}_s = 0.25$, it quickly descends into low-shear-rate (or quasi-static) regions where the DEM simulations show more rapid interface progression than model predictions.

This discrepancy highlights the continuum model's inability to adequately capture the distinct physics that governs regime extremes. A very recent study by Kumawat~\textit{et~al.}~\cite{kumawat2024transient} also identifies consistent differences between DEM simulations and a continuum model for segregation. The authors attributed discrepancies to dimensional constraints in capturing three-dimensional instability patterns. Our analysis reveals alternative source mechanisms that can lead to differences, which can persist even in geometrically simple systems like single-intruder configurations. In quasi-static conditions ($\langle I \rangle< 0.01$), particle motion is dominated by enduring inter-particle contacts and high local friction. Segregation processes become increasingly influenced by creeping motion, such as diffusive remixing and local structural rearrangements. The model's shear-rate-dependent segregation velocity scaling cannot capture these creeping dynamics, systematically underestimating segregation rates in DEM simulations. In collisional flows ($\langle I \rangle > 0.1$), momentum transfer occurs through rapid binary collisions and large velocity fluctuations - a highly agitated regime where local fluctuations enhance diffusive remixing. The continuum model’s formulation essentially assumes a linear response to local rheology, neglecting key nonlinear effects (such as nonlocal stress transmission via force chains in quasi-static zones, shear-enhanced diffusion in collisional flows, jamming and other compressibility-related phenomena) that become significant outside the intermediate $\langle I \rangle$ regime. As a result, the model’s fixed scaling law cannot adapt to these mechanisms, which explains the marked increase in the difference $\Delta\phi_{s} = \phi_{s,\rm CM}(t^{*},z^{*})-\phi_{s,\rm DEM}(t^{*},z^{*})$ shown in the top panels of Fig.~\ref{fig:interface_gradients}. This growing discrepancy becomes particularly evident when the dynamic interface between small and large particles reaches the low-$\langle I \rangle$ deep zone for $\bar{\phi}_{s} = 0.25$, and the high-$\langle I \rangle$ shallow zone for $\bar{\phi}_{s} = 0.75$. The vertical position of this interface—defined by the location of the maximum vertical gradient in $\phi_{s}$ (see bottom panels in Fig.~\ref{fig:interface_gradients}) is especially relevant, as it marks the region where segregation and diffusive-remixing processes are most active.

\section{\label{conclusion}Conclusion and outlook}

This study investigated the dynamics of particle size segregation in granular flows, with a focus on the behavior of segregation velocity under varying rheological conditions. Using DEM numerical experiments, we analyzed the segregation of single intruders and extended the findings to bidisperse mixtures. The results demonstrate that within a moderate inertial number range (approximately $0.01<\langle I \rangle<0.1$), the segregation velocity exhibits a linear dependence on the local rheological conditions, roughly consistent with scaling relations suggested by earlier investigations of segregation in moderately sheared granular flows~\cite{fry2018effect, trewhela2021experimental, jing2022drag, fan2014modelling, chassagne2020discrete,bancroft2021drag}. This scaling highlights the utility of continuum models in capturing segregation dynamics under these conditions. In addition, this study proposed a functional form to capture the influence of the size ratio on segregation velocity, $\mathcal{F}(R)$, which effectively incorporates the physical constraint and observation in the simulations. 

However, deviations from this linear dependence were observed in both the upper and lower regions of the simulation domain, where the segregation velocity becomes less sensitive to changes in $\langle \dot{\gamma} \rangle/\langle p \rangle$ or inertial number $\langle I \rangle$. This shows the limitation of many of the existing scalings for the segregation velocity magnitude when the local inertial number $\langle I \rangle$ falls outside the moderate range. The latter suggests that in these regimes, other mechanisms, such as local packing configurations, diffusion, or collisional dissipation, may play a more dominant role. When applied to bidisperse mixtures, the continuum model effectively captures the overall trends of the segregation fluxes, but underestimates the final segregation states, particularly for systems with a higher fraction of small particles. This trend mirrors the results observed for the cases of small and large intruders, where weak shear rate conditions pose challenges to improve the model's predictions.

We stress the need to refine segregation velocity models to account for regime-dependent behaviors observed under widely varying granular conditions. Additionally, incorporating local packing configurations and investigating their statistical characteristics and their influence on size segregation could provide deeper insights into the origins of the deviations and noise observed in the intruder trajectory and segregation velocity. Future work could also explore these refinements and extend the framework to polydisperse systems, thereby enhancing its applicability to natural and industrial granular flows.\\

\section*{\textbf{ACKNOWLEDGMENT}}
TZ gratefully acknowledges financial support from China Scholarship Council (No. 202306090297). TZ and XL were supported by the China NSF fund 51609179. TT was supported by the ``Agencia National de Investigacion y Desarrollo (ANID)'' through FONDECYT Iniciación Project 11240630. Additionally, DN and HNU were supported by the University of Pennsylvania start-up grant. \\


\noindent{\sf\textbf{AUTHOR DECLARATION}}\\
\noindent{\sf\textbf{Conflict of interest}}

The authors have no conflicts to disclose.\\

\noindent{\sf\textbf{Author Contributions}}\\
\textbf{Tianxiong Zhao}: Conceptualization; Data curation (lead); Formal analysis (lead); Funding acquisition; Investigation; Methodology (lead); Writing - original draft (lead). \textbf{Daisuke Noto}: Supervision; Writing  - original draft. \textbf{Xia Li}: Supervision; Funding acquisition. \textbf{Tom\'as Trewhela}: Conceptualization; Investigation; Formal analysis; Methodology; Supervision; Funding acquisition;  Writing  - original draft. \textbf{Hugo N. Ulloa}: Conceptualization; Investigation; Funding acquisition; Supervision; Writing  - original draft. \\

\noindent{\sf\textbf{DATA AVAILABILITY}}

The data that support the findings of this study are available in \citet{Zhao_zenodo_2025}.

\begin{appendix}
\renewcommand{\thefigure}{\Alph{section}\arabic{figure}}
\appendix

\section{\textcolor{black}{Comparison with \citet{ferdowsi2017river}}}
\label{app:validation}
\setcounter{figure}{0} 

\begin{figure}[h!]
    \centering
    \includegraphics{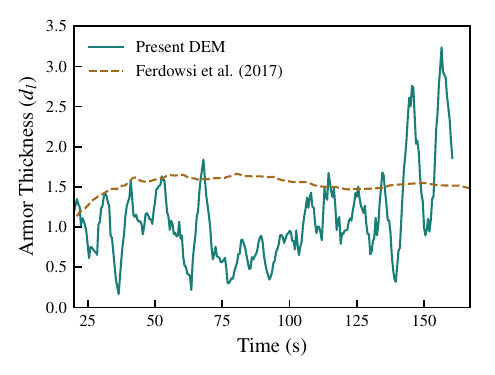}
    \caption{\textcolor{black}{Temporal development of armor thickness. Armor thickness ($z_{1} - z_2$) is determined following \citet{ferdowsi2017river}'s method: $z_{1}$ denotes the top surface where streamwise-averaged large-grain concentration $\phi_l = 0.9$, and $z_2$ is the interface at the minimum $\phi_l$ gradient below the surface.}}
    \label{fig:armor_thickness_validation}
\end{figure}

\textcolor{black}{We benchmark our DEM framework for size-segregation dynamics against the laboratory experiments of \citet{ferdowsi2017river}, who investigated shear-driven segregation and surface armoring in a bidisperse granular system within an annular channel. We replicate their experimental setup using identical particle sizes ($d_s = 1.5$ mm, $d_l = 3.0$ mm), material properties, driving velocity ($u_{\text{top}} = 0.08$ m/s), and boundary conditions.}

\textcolor{black}{Figure~\ref{fig:armor_thickness_validation} presents a quantitative comparison between our DEM simulations and the experimental data for the temporal evolution of the armor layer thickness, normalized by the large particle diameter ($d_l = 3$ mm). Both datasets show similar temporal trends. The observed fluctuations in our results may stem from differences in the coarse-graining spatial resolution}

\section{Background granular flow characterization}
\label{averaging}
\setcounter{figure}{0}
\subsection{Velocity and Shear Rate}

To estimate the background bulk velocity field and shear rate, all instantaneous particle velocities in the streamwise direction $v_x$ - from every timestep and every simulation case - are collected into a concatenated data set. The vertical domain was divided into bins of size $\Delta z_{\text{bin}} = 1.5$~mm, which corresponds to the diameter of the surrounding particles. For a given bin centered at $z_k$, the ensemble average streamwise velocity $\langle {v}_{x} \rangle$ is determined as:
\begin{equation} 
\langle {v}_{x} \rangle|_{z=z_k} = \frac{1}{N_p}\sum_{i=1}^{N_p} v_{x,i}, 
\end{equation}
where $N_p$ is the amount of particles in the bin, and $v_{x,i}$ is the instantaneous streamwise velocity of the particle $i$. Note that this operation is also an implicit lateral averaging. The standard deviation of the streamwise velocity in each bin, denoted by $\sigma_{v_x}$, is also calculated to provide information about the variability of the particle velocity.

The granular flow shear rate $\dot{\gamma}$ is obtained from the laterally averaged streamwise velocity $\overline{v_{x}}$. For a given time $t$, $\overline{v_{x}}|_{z=z_k}$ is calculated by averaging instantaneous particle velocities within each vertical bin, following the same spatial partitioning method. This yields an instantaneous velocity profile for that time step. The shear rate profile at time $t$ is then determined by differentiating $\overline{v_{x}}(t)$ with respect to $z$. For interior bins, the shear rate is calculated as:
\begin{equation}
    \dot{\gamma}|_{z=z_k} = \bigg| \frac{\,\overline{v_{x}}(z_{k+1})-\overline{v_{x}}(z_{k-1})\,}{\,z_{k+1}-z_{k-1}\,} \bigg|.
    \label{eq:shear-rate-profile}
\end{equation}
while boundary bins are calculated using first-order forward or backward differences. The obtained instantaneous shear rate profiles $\dot{\gamma}(z,t)$ are temporally averaged within each simulation case to obtain time-averaged profiles. These profiles are subsequently averaged over all simulation cases, resulting in the ensemble-averaged shear rate $\langle \dot{\gamma} \rangle$.

\subsection{Volume Fraction}

Similar to the procedure used for estimating the bulk velocity field and shear rate, the vertical domain was divided into vertical bins of size $\Delta z_{\text{bin}} = 1.5$~mm. The volume fraction profile, $\Phi$, was then calculated using a two-stage averaging process. First, for each simulated case and every timestep $n$, we computed the instantaneous volume fraction $\Phi(z_k,t)$ in bin $k$ (centered at $z_k$), and performed time averaging:
\begin{equation}
        \langle \Phi(z_k) \rangle_{\mathrm{time}} = \frac{1}{N_t} \sum_{i=1}^{N_t} \Phi(z_k, t)\,,
\end{equation}
where $N_t$ is the number of timesteps in the averaging time window.

Second, the results from different cases were ensemble-averaged to reduce case-specific variability:
\begin{equation}
        \langle \Phi \rangle |_{z=z_k}  = \frac{1}{N_c} \sum_{j=1}^{N_c} \langle \Phi(z_k) \rangle_{\mathrm{time},j}\,,
\end{equation}
where the $N_c$ is the number of cases. 

\subsection{Stress components}

The vertical distribution of the stress components was determined by the analysis of the forces between particles.


As mentioned earlier, the vertical domain $[z_{\text{b}}, z_{\text{t}}]$ was discretized in bins of size $\Delta z_{\text{step}}=1.5$~mm. For a given plane at $z_k$, the particles are divided into two groups: those above the plane ($z_p > z_k$) and those below the plane ($z_p < z_k$). The net force $F(z_k)$ at $z_k$ is defined as the sum of the contact forces acting between particles in the lower and upper groups:
\begin{equation}
    F(z_k) = \sum_{i} f_i,
    \label{eq:force_sum}
\end{equation}
where $f_i$ is the contact force between $i$th pair of particles. From this force balance, the shear stress $\tau_{zx}=\tau$ and the normal stress $p$ are calculated as
\begin{equation}
    \tau(z_k) = \frac{F_x(z_k)}{A} \quad \text{and} \quad p(z_k) = \frac{F_z(z_k)}{A},
    \label{eq:stress}
\end{equation}
where $A$ is the horizontal cross-sectional area with unit normal vector $\bm{k}$ in the vertical $z$-direction.

For each simulation case, the instantaneous values for the stress components at $z_k$ are computed at every time-step $t_{n}$ within the analysis window. The time-averaged stresses are then obtained by
\begin{equation}
     \langle \tau(z_k) \rangle_{\mathrm{time}} = \frac{1}{N_t} \sum_{t=t_{\mathrm{start}}}^{t_{\mathrm{end}}} \tau(z_k, t)\,,
\end{equation}
\begin{equation}
     \langle p(z_k) \rangle_{\mathrm{time}} = \frac{1}{N_t} \sum_{t=t_{\mathrm{start}}}^{t_{\mathrm{end}}} p(z_k, t),
\end{equation}
where $N_t$ is the number of timesteps in each simulation.

Thus, considering all the cases, we estimate the ensemble-averaged stress profiles as follows:
\begin{equation}
     \langle \tau \rangle|_{z=z_k} = \frac{1}{N_c} \sum_{j=1}^{N_c} \langle \tau(z_k) \rangle_{\mathrm{time}, j}\,,
\end{equation}
\begin{equation}
     \langle p \rangle |_{z=z_k} = \frac{1}{N_c} \sum_{j=1}^{N_c} \langle p(z_k) \rangle_{\mathrm{time}, j}\,,
\end{equation}
where $N_c$ is the number of simulation cases.

\section{Intruder trajectory}
\setcounter{figure}{0}

\subsection{Trajectory estimation and velocity calculation}

We process the intruder trajectory data to estimate its segregation velocity $w_\text{seg}$; its vertical position is recorded as a time series, $z_{i}(t)$. In order to represent how fast the intruder's segregates, its trajectory is divided into a series of \emph{checkpoints} `$n$' along the vertical $z$-axis. These checkpoints are spaced at a fixed vertical displacement, equal to the mean diameter of the surrounding particles, $1.5$~mm. This displacement selection ensures that the length scale is large enough to be within the superdiffusive regime \cite{van2015underlying}.

For each adjacent pair of checkpoints, the corresponding \emph{pass time} is determined by identifying the moment when the intruder crosses the checkpoint. The velocity between two consecutive checkpoints is then computed as:
\begin{equation}\label{eq:seg_vel_estimation}
    w_{\text{seg}}(z_{i,n+1/2}) \approx \frac{z_{i}(t_{n+1}) - z_{i}(t_n)}{t_{n+1} - t_n},
\end{equation}
with the representative vertical position for the segment (between the checkpoints $n$ and $n+1$) denoted by $z_{i,n+1/2} = [z_i(t_n)+z_i(t_{n+1})]/2$.

To determine the relationship between the local intruder segregation velocity and the local stress conditions, we scaled the original velocity data (as a function of $z$) using the vertical distribution of the shear rate $\dot\gamma$ and the stress components ($\tau$, $p$). From the surrounding particle information introduced in the previous section, the local effective normal viscosity, $p/\dot\gamma$, is obtained as a function of $z$. The segregation velocity $w_\text{seg}$ can then be scaled as a function of $\dot\gamma/p$

\subsection{Parameter fitting procedure}
\label{FRfitting}

The magnitude of segregation velocity is modeled as follows \cite{trewhela2021experimental}: 
\begin{equation}
    f_{sl} = \left(\frac{\dot{\gamma}\,\rho\,g\,\bar{d}^2}{p}\right)\mathcal{F}(R),
\end{equation}
where $\rho$ is the particles intrinsic density, $g$ is the gravitational acceleration, $\bar{d} = \phi_s d_s+\phi_ld_l$ is the concentration averaged particle diameter, $p/\gamma$ is the local effective normal viscosity, $R$ is the size ratio of the intruder, and $\mathcal{F}(R)$ is assumed to scale as
\begin{equation}
    \mathcal{F}(R) = A_1\,(R-1)^{B_1},
\end{equation}
where $A_1$ and $B_1$ are fitting parameters. Substituting $\mathcal{F}(R)$ into the model, we obtain
\begin{equation}
    f_{sl} =  \left(\frac{\dot{\gamma}\,\rho\,g\,\bar{d}^2}{p}\right)\,A_1\,(R-1)^{B_1}.
\end{equation}

In the case of a single large intruder, $f_{sl}$ equal to the magnitude of the intruder particle, $|w_{\text{seg}}|$. To determine the parameters $A_1$ and $B_1$, we aggregate the estimated values of $|w_{\mathrm{seg}}|$ and $\gamma/p$ from each simulated cases with different size ratio $R$, and perform a non-linear least-squares fit to minimize the residual
\begin{equation}
    \sum_{i} \left[|w_{\mathrm{seg},i}| - \left\{\left(\frac{\dot{\gamma}\,\rho\,g\,\bar{d}^2_s}{p}\right)\,A_1\,(R-1)^{B_1}\right\}_{i}\right]^2,
\end{equation}
where the index $i$ runs over all data points from all cases.

\section{Jing \textit{et al.} (2022) Scaling Analysis}\label{scaling_Jing}
\setcounter{figure}{0}

\setcounter{figure}{0} 
\renewcommand{\thefigure}{D\arabic{figure}} 

\begin{figure}[htbp]
    \centering
    \includegraphics{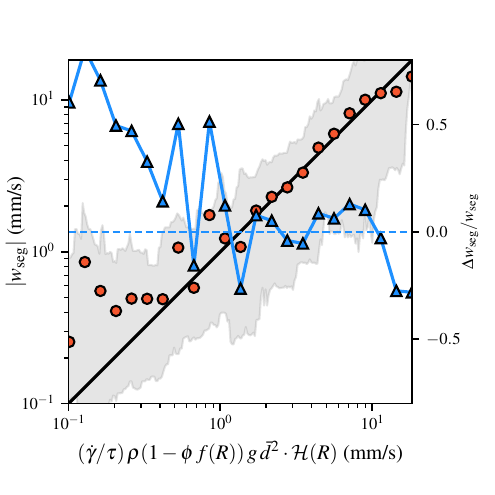}
    \caption{DEM-measured segregation velocity ($w_{\text{seg}}$) plotted against the scaling variable from Jing \textit{et al.}'s framework}
    \label{fig:D1}
\end{figure}

The segregation velocity scaling relationship based on $\tau/\dot{\gamma}$ introduced by ~\citet{jing2022drag} was analyzed using the expression:
\begin{equation}
    |w_{\text{seg}}| = \left(\frac{\dot{\gamma}}{\tau}\right)\rho[1-\phi f(R)]\,g\,\bar{d}^2\,\mathcal{H}(R),
\end{equation}
where $\mathcal{H}(R) = A_3(R-1)^{B_3}$ was determined through nonlinear least-squares fitting to our DEM dataset. The parameter optimization yielded $A_3 = 3.27 \pm 0.11$ and $B_3 = 0.18 \pm 0.05$. Results are shown in Fig.~\ref{fig:D1}. Relative errors were calculated as $\Delta w_{\text{seg}}/w_{\text{seg}} = (w_{\text{seg,DEM}} - w_{\text{seg,model}})/w_{\text{seg,DEM}}$. The shaded gray band shows the 10th-90th percentile ranges from DEM simulations, with orange circles indicating binned mean values. Blue triangles represent relative error $\Delta w_{\text{seg}}/w_{\text{seg}}$ between measurements and model predictions. The dashed line shows ideal agreement. As discussed in the main text, this scaling captures the general segregation behavior and exhibits a similar trend and performance to the scaling proposed by Trewhela \textit{et al.}~\cite{trewhela2021experimental}.

\section{Mixture segregation analysis}\label{DEM mixture analysis}

\setcounter{figure}{0} 
\renewcommand{\thefigure}{E\arabic{figure}} 

 \begin{figure*}[t!]\centering\includegraphics{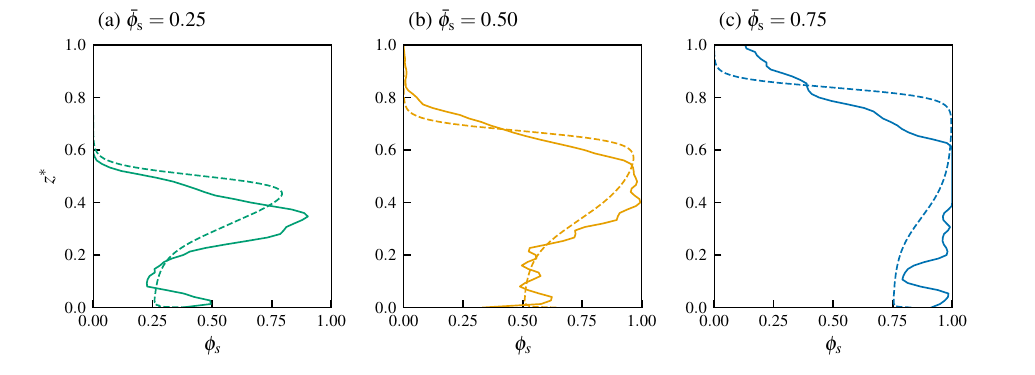}
    \caption{\textcolor{black}{Vertical distribution of small particle concentration $\phi_s$ at the final simulation state for three mixture cases: (a) $\bar{\phi}_s = 25\%$, (b) $\bar{\phi}_s = 50\%$, and (c) $\bar{\phi}_s = 75\%$. Solid lines show DEM results; dashed lines show continuum model (CM) predictions.}}
    \label{fig:app:E1}
\end{figure*}

From the particle position data, we calculate the volume fraction $\phi_{r,k}$ of two particle types ($r \in \{\text{small}, \text{large}\}$) in bin $k$ (centered at $z_k$). The vertical domain $[z_{\text{low}}, z_{\text{high}}]$ is divided into bins with:  

\begin{itemize}  
    \item Bin size: $\Delta z_{\text{bin}} = 3\ \text{mm}$, equal to the mean large particle diameter $\bar{d}_l = 3$~mm.  
    \item Bin centers: $z_k = z_{\text{low}} + n \cdot z_{\text{step}}$, $n \in \mathbb{N}$, where $z_\text{step}=3$~mm
    \item Bin boundaries: $z_k - \frac{z_{\text{bin}}}{2} \leq z < z_k + \frac{z_{\text{bin}}}{2}$.  
\end{itemize}

For particle $i$ (radius $r_i$, center $z_i$) and bin $k$, the overlap height $h_{i,k}$ and volume $V_{i,k}$ are:  
\begin{align*}  
    h_{i,k} &= \max\left\{0, \min\left(z_i + r_i, z_k + \frac{z_{\text{bin}}}{2}\right) - \max\left(z_i - r_i, z_k - \frac{z_{\text{bin}}}{2}\right)\right\}, \\  
    V_{i,k} &= \frac{\pi h_{i,k}^2}{3}(3r_i - h_{i,k}) \quad \text{(spherical cap volume)}.  
\end{align*}  
The volume fraction of the $r$ particle species in bin $k$ is:  
\begin{equation}  
    \phi_{r,k} = \frac{\sum_{i \in \text{type }r} V_{i,k}}{\sum_{i \in \text{all particles}} V_{i,k}}.  
\end{equation}  

The obtained volume fraction $\phi_{r}$ obtained from DEM simulations is plotted as a function of the non-dimensional height $z^{*}$ and non-dimensional time $t^{*}$ in Fig.~\ref{fig6}. \textcolor{black}{Figure~\ref{fig:app:E1} shows the vertical distribution of small particle concentration $\phi_s$ at the final simulation state for three mixture cases. These profiles exhibit a S-shaped form with an exponential decay of $\phi_s$ in fully segregated regions, consistent with expected steady-state segregation patterns\cite{thornton2012modeling, wiederseiner2011experimental}.}

\end{appendix}

\section*{\textbf{REFERENCES}}

\bibliography{ref}

\begin{thebibliography}{63}%
\makeatletter
\providecommand \@ifxundefined [1]{%
 \@ifx{#1\undefined}
}%
\providecommand \@ifnum [1]{%
 \ifnum #1\expandafter \@firstoftwo
 \else \expandafter \@secondoftwo
 \fi
}%
\providecommand \@ifx [1]{%
 \ifx #1\expandafter \@firstoftwo
 \else \expandafter \@secondoftwo
 \fi
}%
\providecommand \natexlab [1]{#1}%
\providecommand \enquote  [1]{``#1''}%
\providecommand \bibnamefont  [1]{#1}%
\providecommand \bibfnamefont [1]{#1}%
\providecommand \citenamefont [1]{#1}%
\providecommand \href@noop [0]{\@secondoftwo}%
\providecommand \href [0]{\begingroup \@sanitize@url \@href}%
\providecommand \@href[1]{\@@startlink{#1}\@@href}%
\providecommand \@@href[1]{\endgroup#1\@@endlink}%
\providecommand \@sanitize@url [0]{\catcode `\\12\catcode `\$12\catcode
  `\&12\catcode `\#12\catcode `\^12\catcode `\_12\catcode `\%12\relax}%
\providecommand \@@startlink[1]{}%
\providecommand \@@endlink[0]{}%
\providecommand \url  [0]{\begingroup\@sanitize@url \@url }%
\providecommand \@url [1]{\endgroup\@href {#1}{\urlprefix }}%
\providecommand \urlprefix  [0]{URL }%
\providecommand \Eprint [0]{\href }%
\providecommand \doibase [0]{http://dx.doi.org/}%
\providecommand \selectlanguage [0]{\@gobble}%
\providecommand \bibinfo  [0]{\@secondoftwo}%
\providecommand \bibfield  [0]{\@secondoftwo}%
\providecommand \translation [1]{[#1]}%
\providecommand \BibitemOpen [0]{}%
\providecommand \bibitemStop [0]{}%
\providecommand \bibitemNoStop [0]{.\EOS\space}%
\providecommand \EOS [0]{\spacefactor3000\relax}%
\providecommand \BibitemShut  [1]{\csname bibitem#1\endcsname}%
\let\auto@bib@innerbib\@empty
\bibitem [{\citenamefont {Trewhela}, \citenamefont {Ancey},\ and\ \citenamefont
  {Gray}(2021)}]{trewhela2021experimental}%
  \BibitemOpen
  \bibfield  {author} {\bibinfo {author} {\bibfnamefont {T.}~\bibnamefont
  {Trewhela}}, \bibinfo {author} {\bibfnamefont {C.}~\bibnamefont {Ancey}}, \
  and\ \bibinfo {author} {\bibfnamefont {J.}~\bibnamefont {Gray}},\ }\bibfield
  {title} {\enquote {\bibinfo {title} {An experimental scaling law for
  particle-size segregation in dense granular flows},}\ }\href {\doibase
  https://doi.org/10.1017/jfm.2021.227} {\bibfield  {journal} {\bibinfo
  {journal} {J. Fluid Mech.}\ }\textbf {\bibinfo {volume} {916}},\ \bibinfo
  {pages} {A55} (\bibinfo {year} {2021})}\BibitemShut {NoStop}%
\bibitem [{\citenamefont {Jing}\ \emph {et~al.}(2021)\citenamefont {Jing},
  \citenamefont {Ottino}, \citenamefont {Lueptow},\ and\ \citenamefont
  {Umbanhowar}}]{jing2021unified}%
  \BibitemOpen
  \bibfield  {author} {\bibinfo {author} {\bibfnamefont {L.}~\bibnamefont
  {Jing}}, \bibinfo {author} {\bibfnamefont {J.~M.}\ \bibnamefont {Ottino}},
  \bibinfo {author} {\bibfnamefont {R.~M.}\ \bibnamefont {Lueptow}}, \ and\
  \bibinfo {author} {\bibfnamefont {P.~B.}\ \bibnamefont {Umbanhowar}},\
  }\bibfield  {title} {\enquote {\bibinfo {title} {A unified description of
  gravity-and kinematics-induced segregation forces in dense granular flows},}\
  }\href {\doibase https://doi.org/10.1017/jfm.2021.688} {\bibfield  {journal}
  {\bibinfo  {journal} {J. Fluid Mech}\ }\textbf {\bibinfo {volume} {925}},\
  \bibinfo {pages} {A29} (\bibinfo {year} {2021})}\BibitemShut {NoStop}%
\bibitem [{\citenamefont {Rosato}\ \emph {et~al.}(1987)\citenamefont {Rosato},
  \citenamefont {Strandburg}, \citenamefont {Prinz},\ and\ \citenamefont
  {Swendsen}}]{rosato1987brazil}%
  \BibitemOpen
  \bibfield  {author} {\bibinfo {author} {\bibfnamefont {A.}~\bibnamefont
  {Rosato}}, \bibinfo {author} {\bibfnamefont {K.~J.}\ \bibnamefont
  {Strandburg}}, \bibinfo {author} {\bibfnamefont {F.}~\bibnamefont {Prinz}}, \
  and\ \bibinfo {author} {\bibfnamefont {R.~H.}\ \bibnamefont {Swendsen}},\
  }\bibfield  {title} {\enquote {\bibinfo {title} {Why the brazil nuts are on
  top: Size segregation of particulate matter by shaking},}\ }\href {\doibase
  https://doi.org/10.1103/PhysRevLett.58.1038} {\bibfield  {journal} {\bibinfo
  {journal} {Phys. Rev. Lett.}\ }\textbf {\bibinfo {volume} {58}},\ \bibinfo
  {pages} {1038} (\bibinfo {year} {1987})}\BibitemShut {NoStop}%
\bibitem [{\citenamefont {Savage}\ and\ \citenamefont
  {Lun}(1988)}]{savage1988particle}%
  \BibitemOpen
  \bibfield  {author} {\bibinfo {author} {\bibfnamefont {S.}~\bibnamefont
  {Savage}}\ and\ \bibinfo {author} {\bibfnamefont {C.}~\bibnamefont {Lun}},\
  }\bibfield  {title} {\enquote {\bibinfo {title} {Particle size segregation in
  inclined chute flow of dry cohesionless granular solids},}\ }\href {\doibase
  https://doi.org/10.1017/S002211208800103X} {\bibfield  {journal} {\bibinfo
  {journal} {J, Fluid Mech.}\ }\textbf {\bibinfo {volume} {189}},\ \bibinfo
  {pages} {311--335} (\bibinfo {year} {1988})}\BibitemShut {NoStop}%
\bibitem [{\citenamefont {Duran}, \citenamefont {Rajchenbach},\ and\
  \citenamefont {Cl{\'e}ment}(1993)}]{duran1993arching}%
  \BibitemOpen
  \bibfield  {author} {\bibinfo {author} {\bibfnamefont {J.}~\bibnamefont
  {Duran}}, \bibinfo {author} {\bibfnamefont {J.}~\bibnamefont {Rajchenbach}},
  \ and\ \bibinfo {author} {\bibfnamefont {E.}~\bibnamefont {Cl{\'e}ment}},\
  }\bibfield  {title} {\enquote {\bibinfo {title} {Arching effect model for
  particle size segregation},}\ }\href {\doibase
  https://doi.org/10.1103/PhysRevLett.70.2431} {\bibfield  {journal} {\bibinfo
  {journal} {Phys. Rev. Lett.}\ }\textbf {\bibinfo {volume} {70}},\ \bibinfo
  {pages} {2431} (\bibinfo {year} {1993})}\BibitemShut {NoStop}%
\bibitem [{\citenamefont {Knight}, \citenamefont {Jaeger},\ and\ \citenamefont
  {Nagel}(1993)}]{knight1993vibration}%
  \BibitemOpen
  \bibfield  {author} {\bibinfo {author} {\bibfnamefont {J.~B.}\ \bibnamefont
  {Knight}}, \bibinfo {author} {\bibfnamefont {H.~M.}\ \bibnamefont {Jaeger}},
  \ and\ \bibinfo {author} {\bibfnamefont {S.~R.}\ \bibnamefont {Nagel}},\
  }\bibfield  {title} {\enquote {\bibinfo {title} {Vibration-induced size
  separation in granular media: The convection connection},}\ }\href {\doibase
  https://doi.org/10.1103/PhysRevLett.70.3728} {\bibfield  {journal} {\bibinfo
  {journal} {Phys. Rev. Lett.}\ }\textbf {\bibinfo {volume} {70}},\ \bibinfo
  {pages} {3728} (\bibinfo {year} {1993})}\BibitemShut {NoStop}%
\bibitem [{\citenamefont {Ottino}\ and\ \citenamefont
  {Khakhar}(2000)}]{ottino2000mixing}%
  \BibitemOpen
  \bibfield  {author} {\bibinfo {author} {\bibfnamefont {J.~M.}\ \bibnamefont
  {Ottino}}\ and\ \bibinfo {author} {\bibfnamefont {D.~V.}\ \bibnamefont
  {Khakhar}},\ }\bibfield  {title} {\enquote {\bibinfo {title} {Mixing and
  segregation of granular materials},}\ }\href {\doibase
  https://doi.org/10.1146/annurev.fluid.32.1.55} {\bibfield  {journal}
  {\bibinfo  {journal} {Annu. Rev. Fluid Mech..}\ }\textbf {\bibinfo {volume}
  {32}},\ \bibinfo {pages} {55--91} (\bibinfo {year} {2000})}\BibitemShut
  {NoStop}%
\bibitem [{\citenamefont {van~der Vaart}\ \emph {et~al.}(2015)\citenamefont
  {van~der Vaart}, \citenamefont {Gajjar}, \citenamefont {Epely-Chauvin},
  \citenamefont {Andreini}, \citenamefont {Gray},\ and\ \citenamefont
  {Ancey}}]{van2015underlying}%
  \BibitemOpen
  \bibfield  {author} {\bibinfo {author} {\bibfnamefont {K.}~\bibnamefont
  {van~der Vaart}}, \bibinfo {author} {\bibfnamefont {P.}~\bibnamefont
  {Gajjar}}, \bibinfo {author} {\bibfnamefont {G.}~\bibnamefont
  {Epely-Chauvin}}, \bibinfo {author} {\bibfnamefont {N.}~\bibnamefont
  {Andreini}}, \bibinfo {author} {\bibfnamefont {J.}~\bibnamefont {Gray}}, \
  and\ \bibinfo {author} {\bibfnamefont {C.}~\bibnamefont {Ancey}},\ }\bibfield
   {title} {\enquote {\bibinfo {title} {Underlying asymmetry within particle
  size segregation},}\ }\href {\doibase
  https://doi.org/10.1103/PhysRevLett.114.238001} {\bibfield  {journal}
  {\bibinfo  {journal} {Phys. Rev. Lett.}\ }\textbf {\bibinfo {volume} {114}},\
  \bibinfo {pages} {238001} (\bibinfo {year} {2015})}\BibitemShut {NoStop}%
\bibitem [{\citenamefont {Gray}(2018)}]{gray2018particle}%
  \BibitemOpen
  \bibfield  {author} {\bibinfo {author} {\bibfnamefont {J.~M. N.~T.}\
  \bibnamefont {Gray}},\ }\bibfield  {title} {\enquote {\bibinfo {title}
  {Particle segregation in dense granular flows},}\ }\href {\doibase
  https://doi.org/10.1146/annurev-fluid-122316-045201} {\bibfield  {journal}
  {\bibinfo  {journal} {Ann. Rev. Fluid Mech.}\ }\textbf {\bibinfo {volume}
  {50}},\ \bibinfo {pages} {407--433} (\bibinfo {year} {2018})}\BibitemShut
  {NoStop}%
\bibitem [{\citenamefont {Umbanhowar}, \citenamefont {Lueptow},\ and\
  \citenamefont {Ottino}(2019)}]{umbanhowar2019review}%
  \BibitemOpen
  \bibfield  {author} {\bibinfo {author} {\bibfnamefont {P.~B.}\ \bibnamefont
  {Umbanhowar}}, \bibinfo {author} {\bibfnamefont {R.~M.}\ \bibnamefont
  {Lueptow}}, \ and\ \bibinfo {author} {\bibfnamefont {J.~M.}\ \bibnamefont
  {Ottino}},\ }\bibfield  {title} {\enquote {\bibinfo {title} {Modeling
  segregation in granular flows},}\ }\href {\doibase
  https://doi.org/10.1146/annurev-chembioeng-060718-030122} {\bibfield
  {journal} {\bibinfo  {journal} {Ann. Rev. Chem. Biomol. Eng.}\ }\textbf
  {\bibinfo {volume} {10}},\ \bibinfo {pages} {129--153} (\bibinfo {year}
  {2019})}\BibitemShut {NoStop}%
\bibitem [{\citenamefont {C{\'u}{\~n}ez}, \citenamefont {Patel},\ and\
  \citenamefont {Glade}(2024)}]{cunez2024particle}%
  \BibitemOpen
  \bibfield  {author} {\bibinfo {author} {\bibfnamefont {F.~D.}\ \bibnamefont
  {C{\'u}{\~n}ez}}, \bibinfo {author} {\bibfnamefont {D.}~\bibnamefont
  {Patel}}, \ and\ \bibinfo {author} {\bibfnamefont {R.~C.}\ \bibnamefont
  {Glade}},\ }\bibfield  {title} {\enquote {\bibinfo {title} {How particle
  shape affects granular segregation in industrial and geophysical flows},}\
  }\href {\doibase https://doi.org/10.1073/pnas.2307061121} {\bibfield
  {journal} {\bibinfo  {journal} {Proc. Natl. Acad. Sci.}\ }\textbf {\bibinfo
  {volume} {121}},\ \bibinfo {pages} {e2307061121} (\bibinfo {year}
  {2024})}\BibitemShut {NoStop}%
\bibitem [{\citenamefont {F{\'e}lix}\ and\ \citenamefont
  {Thomas}(2004)}]{felix2004relation}%
  \BibitemOpen
  \bibfield  {author} {\bibinfo {author} {\bibfnamefont {G.}~\bibnamefont
  {F{\'e}lix}}\ and\ \bibinfo {author} {\bibfnamefont {N.}~\bibnamefont
  {Thomas}},\ }\bibfield  {title} {\enquote {\bibinfo {title} {Relation between
  dry granular flow regimes and morphology of deposits: formation of lev{\'e}es
  in pyroclastic deposits},}\ }\href {\doibase
  https://doi.org/10.1016/S0012-821X(04)00111-6} {\bibfield  {journal}
  {\bibinfo  {journal} {Earth and Planet. Sci. Lett.}\ }\textbf {\bibinfo
  {volume} {221}},\ \bibinfo {pages} {197--213} (\bibinfo {year}
  {2004})}\BibitemShut {NoStop}%
\bibitem [{\citenamefont {Kleinhans}(2004)}]{kleinhans2004sorting}%
  \BibitemOpen
  \bibfield  {author} {\bibinfo {author} {\bibfnamefont {M.~G.}\ \bibnamefont
  {Kleinhans}},\ }\bibfield  {title} {\enquote {\bibinfo {title} {Sorting in
  grain flows at the lee side of dunes},}\ }\href {\doibase
  https://doi.org/10.1016/S0012-8252(03)00081-3} {\bibfield  {journal}
  {\bibinfo  {journal} {Earth-Sci. Rev.}\ }\textbf {\bibinfo {volume} {65}},\
  \bibinfo {pages} {75--102} (\bibinfo {year} {2004})}\BibitemShut {NoStop}%
\bibitem [{\citenamefont {Johnson}\ \emph {et~al.}(2012)\citenamefont
  {Johnson}, \citenamefont {Kokelaar}, \citenamefont {Iverson}, \citenamefont
  {Logan}, \citenamefont {LaHusen},\ and\ \citenamefont
  {Gray}}]{johnson2012grain}%
  \BibitemOpen
  \bibfield  {author} {\bibinfo {author} {\bibfnamefont {C.~G.}\ \bibnamefont
  {Johnson}}, \bibinfo {author} {\bibfnamefont {B.~P.}\ \bibnamefont
  {Kokelaar}}, \bibinfo {author} {\bibfnamefont {R.~M.}\ \bibnamefont
  {Iverson}}, \bibinfo {author} {\bibfnamefont {M.}~\bibnamefont {Logan}},
  \bibinfo {author} {\bibfnamefont {R.~G.}\ \bibnamefont {LaHusen}}, \ and\
  \bibinfo {author} {\bibfnamefont {J.~M. N.~T.}\ \bibnamefont {Gray}},\
  }\bibfield  {title} {\enquote {\bibinfo {title} {Grain-size segregation and
  levee formation in geophysical mass flows},}\ }\href {\doibase
  https://doi.org/10.1029/2011JF002185} {\bibfield  {journal} {\bibinfo
  {journal} {J. Geophys. Res. Earth Surf.}\ }\textbf {\bibinfo {volume} {117}}
  (\bibinfo {year} {2012}),\ https://doi.org/10.1029/2011JF002185}\BibitemShut
  {NoStop}%
\bibitem [{\citenamefont {Kokelaar}\ \emph {et~al.}(2014)\citenamefont
  {Kokelaar}, \citenamefont {Graham}, \citenamefont {Gray},\ and\ \citenamefont
  {Vallance}}]{kokelaar2014fine}%
  \BibitemOpen
  \bibfield  {author} {\bibinfo {author} {\bibfnamefont {B.~P.}\ \bibnamefont
  {Kokelaar}}, \bibinfo {author} {\bibfnamefont {R.}~\bibnamefont {Graham}},
  \bibinfo {author} {\bibfnamefont {J.}~\bibnamefont {Gray}}, \ and\ \bibinfo
  {author} {\bibfnamefont {J.~W.}\ \bibnamefont {Vallance}},\ }\bibfield
  {title} {\enquote {\bibinfo {title} {Fine-grained linings of leveed channels
  facilitate runout of granular flows},}\ }\href {\doibase
  https://doi.org/10.1016/j.epsl.2013.10.043} {\bibfield  {journal} {\bibinfo
  {journal} {Earth Planet. Sci. Lett.}\ }\textbf {\bibinfo {volume} {385}},\
  \bibinfo {pages} {172--180} (\bibinfo {year} {2014})}\BibitemShut {NoStop}%
\bibitem [{\citenamefont {De~Haas}\ \emph {et~al.}(2015)\citenamefont
  {De~Haas}, \citenamefont {Braat}, \citenamefont {Leuven}, \citenamefont
  {Lokhorst},\ and\ \citenamefont {Kleinhans}}]{de2015effects}%
  \BibitemOpen
  \bibfield  {author} {\bibinfo {author} {\bibfnamefont {T.}~\bibnamefont
  {De~Haas}}, \bibinfo {author} {\bibfnamefont {L.}~\bibnamefont {Braat}},
  \bibinfo {author} {\bibfnamefont {J.~R. F.~W.}\ \bibnamefont {Leuven}},
  \bibinfo {author} {\bibfnamefont {I.~R.}\ \bibnamefont {Lokhorst}}, \ and\
  \bibinfo {author} {\bibfnamefont {M.~G.}\ \bibnamefont {Kleinhans}},\
  }\bibfield  {title} {\enquote {\bibinfo {title} {Effects of debris flow
  composition on runout, depositional mechanisms, and deposit morphology in
  laboratory experiments},}\ }\href {\doibase
  https://doi.org/10.1002/2015JF003525} {\bibfield  {journal} {\bibinfo
  {journal} {J. Geophys. Res. Earth Surf.}\ }\textbf {\bibinfo {volume}
  {120}},\ \bibinfo {pages} {1949--1972} (\bibinfo {year} {2015})}\BibitemShut
  {NoStop}%
\bibitem [{\citenamefont {Chassagne}\ \emph {et~al.}(2020)\citenamefont
  {Chassagne}, \citenamefont {Maurin}, \citenamefont {Chauchat}, \citenamefont
  {Gray},\ and\ \citenamefont {Frey}}]{chassagne2020discrete}%
  \BibitemOpen
  \bibfield  {author} {\bibinfo {author} {\bibfnamefont {R.}~\bibnamefont
  {Chassagne}}, \bibinfo {author} {\bibfnamefont {R.}~\bibnamefont {Maurin}},
  \bibinfo {author} {\bibfnamefont {J.}~\bibnamefont {Chauchat}}, \bibinfo
  {author} {\bibfnamefont {J.}~\bibnamefont {Gray}}, \ and\ \bibinfo {author}
  {\bibfnamefont {P.}~\bibnamefont {Frey}},\ }\bibfield  {title} {\enquote
  {\bibinfo {title} {Discrete and continuum modelling of grain size segregation
  during bedload transport},}\ }\href {\doibase
  https://doi.org/10.1017/jfm.2020.274} {\bibfield  {journal} {\bibinfo
  {journal} {J. Fluid Mech.}\ }\textbf {\bibinfo {volume} {895}},\ \bibinfo
  {pages} {A30} (\bibinfo {year} {2020})}\BibitemShut {NoStop}%
\bibitem [{\citenamefont {Kostynick}\ \emph {et~al.}(2022)\citenamefont
  {Kostynick}, \citenamefont {Matinpour}, \citenamefont {Pradeep},
  \citenamefont {Haber}, \citenamefont {Sauret}, \citenamefont {Meiburg},
  \citenamefont {Dunne}, \citenamefont {Arratia},\ and\ \citenamefont
  {Jerolmack}}]{kostynick2022rheology}%
  \BibitemOpen
  \bibfield  {author} {\bibinfo {author} {\bibfnamefont {R.}~\bibnamefont
  {Kostynick}}, \bibinfo {author} {\bibfnamefont {H.}~\bibnamefont
  {Matinpour}}, \bibinfo {author} {\bibfnamefont {S.}~\bibnamefont {Pradeep}},
  \bibinfo {author} {\bibfnamefont {S.}~\bibnamefont {Haber}}, \bibinfo
  {author} {\bibfnamefont {A.}~\bibnamefont {Sauret}}, \bibinfo {author}
  {\bibfnamefont {E.}~\bibnamefont {Meiburg}}, \bibinfo {author} {\bibfnamefont
  {T.}~\bibnamefont {Dunne}}, \bibinfo {author} {\bibfnamefont
  {P.}~\bibnamefont {Arratia}}, \ and\ \bibinfo {author} {\bibfnamefont
  {D.}~\bibnamefont {Jerolmack}},\ }\bibfield  {title} {\enquote {\bibinfo
  {title} {Rheology of debris flow materials is controlled by the distance from
  jamming},}\ }\href {\doibase https://doi.org/10.1073/pnas.2209109119}
  {\bibfield  {journal} {\bibinfo  {journal} {Proc. Natl. Acad. Sci.}\ }\textbf
  {\bibinfo {volume} {119}},\ \bibinfo {pages} {e2209109119} (\bibinfo {year}
  {2022})}\BibitemShut {NoStop}%
\bibitem [{\citenamefont {Dedieu}\ \emph {et~al.}(2024)\citenamefont {Dedieu},
  \citenamefont {Rousseau}, \citenamefont {Chauchat},\ and\ \citenamefont
  {Frey}}]{dedieu2024sediment}%
  \BibitemOpen
  \bibfield  {author} {\bibinfo {author} {\bibfnamefont {B.}~\bibnamefont
  {Dedieu}}, \bibinfo {author} {\bibfnamefont {H.}~\bibnamefont {Rousseau}},
  \bibinfo {author} {\bibfnamefont {J.}~\bibnamefont {Chauchat}}, \ and\
  \bibinfo {author} {\bibfnamefont {P.}~\bibnamefont {Frey}},\ }\bibfield
  {title} {\enquote {\bibinfo {title} {Exploring the size ratio impact on an
  intruder segregating in bedload transport},}\ }\href {\doibase
  10.1103/PhysRevFluids.9.104302} {\bibfield  {journal} {\bibinfo  {journal}
  {Phys. Rev. Fluids}\ }\textbf {\bibinfo {volume} {9}},\ \bibinfo {pages}
  {104302} (\bibinfo {year} {2024})}\BibitemShut {NoStop}%
\bibitem [{\citenamefont {Trewhela}\ and\ \citenamefont
  {Ulloa}(2024)}]{Trewhela_Ulloa_2024}%
  \BibitemOpen
  \bibfield  {author} {\bibinfo {author} {\bibfnamefont {T.}~\bibnamefont
  {Trewhela}}\ and\ \bibinfo {author} {\bibfnamefont {H.~N.}\ \bibnamefont
  {Ulloa}},\ }\bibfield  {title} {\enquote {\bibinfo {title} {Energetics of
  particle-size segregation},}\ }\href {\doibase
  https://doi.org/10.1017/jfm.2024.1053} {\bibfield  {journal} {\bibinfo
  {journal} {J. Fluid Mech.}\ }\textbf {\bibinfo {volume} {1000}},\ \bibinfo
  {pages} {A50} (\bibinfo {year} {2024})}\BibitemShut {NoStop}%
\bibitem [{\citenamefont {Makse}\ \emph {et~al.}(1997)\citenamefont {Makse},
  \citenamefont {Havlin}, \citenamefont {King},\ and\ \citenamefont
  {Stanley}}]{makse1997spontaneous}%
  \BibitemOpen
  \bibfield  {author} {\bibinfo {author} {\bibfnamefont {H.~A.}\ \bibnamefont
  {Makse}}, \bibinfo {author} {\bibfnamefont {S.}~\bibnamefont {Havlin}},
  \bibinfo {author} {\bibfnamefont {P.~R.}\ \bibnamefont {King}}, \ and\
  \bibinfo {author} {\bibfnamefont {H.~E.}\ \bibnamefont {Stanley}},\
  }\bibfield  {title} {\enquote {\bibinfo {title} {Spontaneous stratification
  in granular mixtures},}\ }\href {\doibase https://doi.org/10.1038/386379a0}
  {\bibfield  {journal} {\bibinfo  {journal} {Nature}\ }\textbf {\bibinfo
  {volume} {386}},\ \bibinfo {pages} {379--382} (\bibinfo {year}
  {1997})}\BibitemShut {NoStop}%
\bibitem [{\citenamefont {Shinbrot}\ and\ \citenamefont
  {Muzzio}(2000)}]{shinbrot2000nonequilibrium}%
  \BibitemOpen
  \bibfield  {author} {\bibinfo {author} {\bibfnamefont {T.}~\bibnamefont
  {Shinbrot}}\ and\ \bibinfo {author} {\bibfnamefont {F.~J.}\ \bibnamefont
  {Muzzio}},\ }\bibfield  {title} {\enquote {\bibinfo {title} {Nonequilibrium
  patterns in granular mixing and segregation.}}\ }\href {\doibase
  https://doi.org/10.1063/1.883018} {\bibfield  {journal} {\bibinfo  {journal}
  {Physics Today}\ }\textbf {\bibinfo {volume} {53}},\ \bibinfo {pages}
  {25--30} (\bibinfo {year} {2000})}\BibitemShut {NoStop}%
\bibitem [{\citenamefont {Wormsbecker}\ \emph {et~al.}(2005)\citenamefont
  {Wormsbecker}, \citenamefont {Adams}, \citenamefont {Pugsley},\ and\
  \citenamefont {Winters}}]{wormsbecker2005segregation}%
  \BibitemOpen
  \bibfield  {author} {\bibinfo {author} {\bibfnamefont {M.}~\bibnamefont
  {Wormsbecker}}, \bibinfo {author} {\bibfnamefont {A.}~\bibnamefont {Adams}},
  \bibinfo {author} {\bibfnamefont {T.}~\bibnamefont {Pugsley}}, \ and\
  \bibinfo {author} {\bibfnamefont {C.}~\bibnamefont {Winters}},\ }\bibfield
  {title} {\enquote {\bibinfo {title} {Segregation by size difference in a
  conical fluidized bed of pharmaceutical granulate},}\ }\href {\doibase
  https://doi.org/10.1016/j.powtec.2005.02.006} {\bibfield  {journal} {\bibinfo
   {journal} {Powder Technol.}\ }\textbf {\bibinfo {volume} {153}},\ \bibinfo
  {pages} {72--80} (\bibinfo {year} {2005})}\BibitemShut {NoStop}%
\bibitem [{\citenamefont {Guillard}, \citenamefont {Forterre},\ and\
  \citenamefont {Pouliquen}(2016)}]{guillard2016scaling}%
  \BibitemOpen
  \bibfield  {author} {\bibinfo {author} {\bibfnamefont {F.}~\bibnamefont
  {Guillard}}, \bibinfo {author} {\bibfnamefont {Y.}~\bibnamefont {Forterre}},
  \ and\ \bibinfo {author} {\bibfnamefont {O.}~\bibnamefont {Pouliquen}},\
  }\bibfield  {title} {\enquote {\bibinfo {title} {Scaling laws for segregation
  forces in dense sheared granular flows},}\ }\href {\doibase
  https://doi.org/10.1017/jfm.2016.605} {\bibfield  {journal} {\bibinfo
  {journal} {J. Fluid Mech.}\ }\textbf {\bibinfo {volume} {807}},\ \bibinfo
  {pages} {R1} (\bibinfo {year} {2016})}\BibitemShut {NoStop}%
\bibitem [{\citenamefont {Jing}\ \emph {et~al.}(2020)\citenamefont {Jing},
  \citenamefont {Ottino}, \citenamefont {Lueptow},\ and\ \citenamefont
  {Umbanhowar}}]{jing2020rising}%
  \BibitemOpen
  \bibfield  {author} {\bibinfo {author} {\bibfnamefont {L.}~\bibnamefont
  {Jing}}, \bibinfo {author} {\bibfnamefont {J.~M.}\ \bibnamefont {Ottino}},
  \bibinfo {author} {\bibfnamefont {R.~M.}\ \bibnamefont {Lueptow}}, \ and\
  \bibinfo {author} {\bibfnamefont {P.~B.}\ \bibnamefont {Umbanhowar}},\
  }\bibfield  {title} {\enquote {\bibinfo {title} {Rising and sinking intruders
  in dense granular flows},}\ }\href {\doibase
  https://doi.org/10.1103/PhysRevResearch.2.022069} {\bibfield  {journal}
  {\bibinfo  {journal} {Phys. Rev. Res.}\ }\textbf {\bibinfo {volume} {2}},\
  \bibinfo {pages} {022069} (\bibinfo {year} {2020})}\BibitemShut {NoStop}%
\bibitem [{\citenamefont {Bridgwater}, \citenamefont {Foo},\ and\ \citenamefont
  {Stephens}(1985)}]{bridgwater1985particle}%
  \BibitemOpen
  \bibfield  {author} {\bibinfo {author} {\bibfnamefont {J.}~\bibnamefont
  {Bridgwater}}, \bibinfo {author} {\bibfnamefont {W.}~\bibnamefont {Foo}}, \
  and\ \bibinfo {author} {\bibfnamefont {D.}~\bibnamefont {Stephens}},\
  }\bibfield  {title} {\enquote {\bibinfo {title} {Particle mixing and
  segregation in failure zones—theory and experiment},}\ }\href {\doibase
  https://doi.org/10.1016/0032-5910(85)87033-9} {\bibfield  {journal} {\bibinfo
   {journal} {Powder Technol.}\ }\textbf {\bibinfo {volume} {41}},\ \bibinfo
  {pages} {147--158} (\bibinfo {year} {1985})}\BibitemShut {NoStop}%
\bibitem [{\citenamefont {Jing}\ \emph {et~al.}(2022)\citenamefont {Jing},
  \citenamefont {Ottino}, \citenamefont {Umbanhowar},\ and\ \citenamefont
  {Lueptow}}]{jing2022drag}%
  \BibitemOpen
  \bibfield  {author} {\bibinfo {author} {\bibfnamefont {L.}~\bibnamefont
  {Jing}}, \bibinfo {author} {\bibfnamefont {J.~M.}\ \bibnamefont {Ottino}},
  \bibinfo {author} {\bibfnamefont {P.~B.}\ \bibnamefont {Umbanhowar}}, \ and\
  \bibinfo {author} {\bibfnamefont {R.~M.}\ \bibnamefont {Lueptow}},\
  }\bibfield  {title} {\enquote {\bibinfo {title} {Drag force in granular shear
  flows: regimes, scaling laws and implications for segregation},}\ }\href
  {\doibase https://doi.org/10.1017/jfm.2022.706} {\bibfield  {journal}
  {\bibinfo  {journal} {J. Fluid Mech.}\ }\textbf {\bibinfo {volume} {948}},\
  \bibinfo {pages} {A24} (\bibinfo {year} {2022})}\BibitemShut {NoStop}%
\bibitem [{\citenamefont {Duan}\ \emph {et~al.}(2022)\citenamefont {Duan},
  \citenamefont {Jing}, \citenamefont {Umbanhowar}, \citenamefont {Ottino},\
  and\ \citenamefont {Lueptow}}]{duan2022segregation}%
  \BibitemOpen
  \bibfield  {author} {\bibinfo {author} {\bibfnamefont {Y.}~\bibnamefont
  {Duan}}, \bibinfo {author} {\bibfnamefont {L.}~\bibnamefont {Jing}}, \bibinfo
  {author} {\bibfnamefont {P.~B.}\ \bibnamefont {Umbanhowar}}, \bibinfo
  {author} {\bibfnamefont {J.~M.}\ \bibnamefont {Ottino}}, \ and\ \bibinfo
  {author} {\bibfnamefont {R.~M.}\ \bibnamefont {Lueptow}},\ }\bibfield
  {title} {\enquote {\bibinfo {title} {Segregation forces in dense granular
  flows: closing the gap between single intruders and mixtures},}\ }\href
  {\doibase https://doi.org/10.1017/jfm.2022.12} {\bibfield  {journal}
  {\bibinfo  {journal} {J. Fluid Mech.}\ }\textbf {\bibinfo {volume} {935}},\
  \bibinfo {pages} {R1} (\bibinfo {year} {2022})}\BibitemShut {NoStop}%
\bibitem [{\citenamefont {Tripathi}\ and\ \citenamefont
  {Khakhar}(2011)}]{tripathi2011numerical}%
  \BibitemOpen
  \bibfield  {author} {\bibinfo {author} {\bibfnamefont {A.}~\bibnamefont
  {Tripathi}}\ and\ \bibinfo {author} {\bibfnamefont {D.}~\bibnamefont
  {Khakhar}},\ }\bibfield  {title} {\enquote {\bibinfo {title} {Numerical
  simulation of the sedimentation of a sphere in a sheared granular fluid: A
  granular stokes experiment},}\ }\href {\doibase
  https://doi.org/10.1103/PhysRevLett.107.108001} {\bibfield  {journal}
  {\bibinfo  {journal} {Phys. Rev. Lett.}\ }\textbf {\bibinfo {volume} {107}},\
  \bibinfo {pages} {108001} (\bibinfo {year} {2011})}\BibitemShut {NoStop}%
\bibitem [{\citenamefont {Fan}\ \emph {et~al.}(2014)\citenamefont {Fan},
  \citenamefont {Schlick}, \citenamefont {Umbanhowar}, \citenamefont {Ottino},\
  and\ \citenamefont {Lueptow}}]{fan2014modelling}%
  \BibitemOpen
  \bibfield  {author} {\bibinfo {author} {\bibfnamefont {Y.}~\bibnamefont
  {Fan}}, \bibinfo {author} {\bibfnamefont {C.~P.}\ \bibnamefont {Schlick}},
  \bibinfo {author} {\bibfnamefont {P.~B.}\ \bibnamefont {Umbanhowar}},
  \bibinfo {author} {\bibfnamefont {J.~M.}\ \bibnamefont {Ottino}}, \ and\
  \bibinfo {author} {\bibfnamefont {R.~M.}\ \bibnamefont {Lueptow}},\
  }\bibfield  {title} {\enquote {\bibinfo {title} {Modelling size segregation
  of granular materials: the roles of segregation, advection and diffusion},}\
  }\href {\doibase https://doi.org/10.1017/jfm.2013.680} {\bibfield  {journal}
  {\bibinfo  {journal} {J. Fluid Mech.}\ }\textbf {\bibinfo {volume} {741}},\
  \bibinfo {pages} {252--279} (\bibinfo {year} {2014})}\BibitemShut {NoStop}%
\bibitem [{\citenamefont {Fry}\ \emph {et~al.}(2018)\citenamefont {Fry},
  \citenamefont {Umbanhowar}, \citenamefont {Ottino},\ and\ \citenamefont
  {Lueptow}}]{fry2018effect}%
  \BibitemOpen
  \bibfield  {author} {\bibinfo {author} {\bibfnamefont {A.~M.}\ \bibnamefont
  {Fry}}, \bibinfo {author} {\bibfnamefont {P.~B.}\ \bibnamefont {Umbanhowar}},
  \bibinfo {author} {\bibfnamefont {J.~M.}\ \bibnamefont {Ottino}}, \ and\
  \bibinfo {author} {\bibfnamefont {R.~M.}\ \bibnamefont {Lueptow}},\
  }\bibfield  {title} {\enquote {\bibinfo {title} {Effect of pressure on
  segregation in granular shear flows},}\ }\href {\doibase
  https://doi.org/10.1103/PhysRevE.97.062906} {\bibfield  {journal} {\bibinfo
  {journal} {Phys. Rev. E}\ }\textbf {\bibinfo {volume} {97}},\ \bibinfo
  {pages} {062906} (\bibinfo {year} {2018})}\BibitemShut {NoStop}%
\bibitem [{\citenamefont {Frey}\ \emph {et~al.}(2020)\citenamefont {Frey},
  \citenamefont {De~Micheaux}, \citenamefont {Bel}, \citenamefont {Maurin},
  \citenamefont {Rorsman}, \citenamefont {Martin},\ and\ \citenamefont
  {Ducottet}}]{frey2020experiments}%
  \BibitemOpen
  \bibfield  {author} {\bibinfo {author} {\bibfnamefont {P.}~\bibnamefont
  {Frey}}, \bibinfo {author} {\bibfnamefont {H.~L.}\ \bibnamefont
  {De~Micheaux}}, \bibinfo {author} {\bibfnamefont {C.}~\bibnamefont {Bel}},
  \bibinfo {author} {\bibfnamefont {R.}~\bibnamefont {Maurin}}, \bibinfo
  {author} {\bibfnamefont {K.}~\bibnamefont {Rorsman}}, \bibinfo {author}
  {\bibfnamefont {T.}~\bibnamefont {Martin}}, \ and\ \bibinfo {author}
  {\bibfnamefont {C.}~\bibnamefont {Ducottet}},\ }\bibfield  {title} {\enquote
  {\bibinfo {title} {Experiments on grain size segregation in bedload transport
  on a steep slope},}\ }\href {\doibase
  https://doi.org/10.1016/j.advwatres.2019.103478} {\bibfield  {journal}
  {\bibinfo  {journal} {Adv. Water Resour.}\ }\textbf {\bibinfo {volume}
  {136}},\ \bibinfo {pages} {103478} (\bibinfo {year} {2020})}\BibitemShut
  {NoStop}%
\bibitem [{\citenamefont {Trewhela}, \citenamefont {Gray},\ and\ \citenamefont
  {Ancey}(2021)}]{trewhela2021large}%
  \BibitemOpen
  \bibfield  {author} {\bibinfo {author} {\bibfnamefont {T.}~\bibnamefont
  {Trewhela}}, \bibinfo {author} {\bibfnamefont {J.~M. N.~T.}\ \bibnamefont
  {Gray}}, \ and\ \bibinfo {author} {\bibfnamefont {C.}~\bibnamefont {Ancey}},\
  }\bibfield  {title} {\enquote {\bibinfo {title} {Large particle segregation
  in two-dimensional sheared granular flows},}\ }\href {\doibase
  https://doi.org/10.1103/PhysRevFluids.6.054302} {\bibfield  {journal}
  {\bibinfo  {journal} {Phys. Rev. Fluids}\ }\textbf {\bibinfo {volume} {6}},\
  \bibinfo {pages} {054302} (\bibinfo {year} {2021})}\BibitemShut {NoStop}%
\bibitem [{\citenamefont {GDR-MiDi}(2004)}]{gdr2004dense}%
  \BibitemOpen
  \bibfield  {author} {\bibinfo {author} {\bibnamefont {GDR-MiDi}},\ }\bibfield
   {title} {\enquote {\bibinfo {title} {On dense granular flows},}\ }\href
  {\doibase https://doi.org/10.1140/epje/i2003-10153-0} {\bibfield  {journal}
  {\bibinfo  {journal} {Eur. Phys. J. E}\ }\textbf {\bibinfo {volume} {14}},\
  \bibinfo {pages} {341--365} (\bibinfo {year} {2004})}\BibitemShut {NoStop}%
\bibitem [{\citenamefont {Gaume}, \citenamefont {Chambon},\ and\ \citenamefont
  {Naaim}(2011)}]{gaume2011quasistatic}%
  \BibitemOpen
  \bibfield  {author} {\bibinfo {author} {\bibfnamefont {J.}~\bibnamefont
  {Gaume}}, \bibinfo {author} {\bibfnamefont {G.}~\bibnamefont {Chambon}}, \
  and\ \bibinfo {author} {\bibfnamefont {M.}~\bibnamefont {Naaim}},\ }\bibfield
   {title} {\enquote {\bibinfo {title} {Quasistatic to inertial transition in
  granular materials and the role of fluctuations},}\ }\href {\doibase
  https://doi.org/10.1103/PhysRevE.84.051304} {\bibfield  {journal} {\bibinfo
  {journal} {Phys. Rev. E}\ }\textbf {\bibinfo {volume} {84}},\ \bibinfo
  {pages} {051304} (\bibinfo {year} {2011})}\BibitemShut {NoStop}%
\bibitem [{\citenamefont {Houssais}\ \emph {et~al.}(2015)\citenamefont
  {Houssais}, \citenamefont {Ortiz}, \citenamefont {Durian},\ and\
  \citenamefont {Jerolmack}}]{houssais2015onset}%
  \BibitemOpen
  \bibfield  {author} {\bibinfo {author} {\bibfnamefont {M.}~\bibnamefont
  {Houssais}}, \bibinfo {author} {\bibfnamefont {C.~P.}\ \bibnamefont {Ortiz}},
  \bibinfo {author} {\bibfnamefont {D.~J.}\ \bibnamefont {Durian}}, \ and\
  \bibinfo {author} {\bibfnamefont {D.~J.}\ \bibnamefont {Jerolmack}},\
  }\bibfield  {title} {\enquote {\bibinfo {title} {Onset of sediment transport
  is a continuous transition driven by fluid shear and granular creep},}\
  }\href {\doibase https://doi.org/10.1038/ncomms7527} {\bibfield  {journal}
  {\bibinfo  {journal} {Nat. Commun.}\ }\textbf {\bibinfo {volume} {6}},\
  \bibinfo {pages} {6527} (\bibinfo {year} {2015})}\BibitemShut {NoStop}%
\bibitem [{\citenamefont {Gonzalez}, \citenamefont {C{\'u}{\~n}ez},\ and\
  \citenamefont {Franklin}(2023)}]{gonzalez2023bidisperse}%
  \BibitemOpen
  \bibfield  {author} {\bibinfo {author} {\bibfnamefont {J.~O.}\ \bibnamefont
  {Gonzalez}}, \bibinfo {author} {\bibfnamefont {F.~D.}\ \bibnamefont
  {C{\'u}{\~n}ez}}, \ and\ \bibinfo {author} {\bibfnamefont {E.~M.}\
  \bibnamefont {Franklin}},\ }\bibfield  {title} {\enquote {\bibinfo {title}
  {Bidisperse beds sheared by viscous fluids: Grain segregation and bed
  hardening},}\ }\href {\doibase https://doi.org/10.1063/5.0168415} {\bibfield
  {journal} {\bibinfo  {journal} {Phys. Fluids}\ }\textbf {\bibinfo {volume}
  {35}} (\bibinfo {year} {2023}),\
  https://doi.org/10.1063/5.0168415}\BibitemShut {NoStop}%
\bibitem [{\citenamefont {Meng}\ \emph {et~al.}(2024)\citenamefont {Meng},
  \citenamefont {Taylor-Noonan}, \citenamefont {Johnson}, \citenamefont {Take},
  \citenamefont {Bowman},\ and\ \citenamefont {Gray}}]{meng2024granular}%
  \BibitemOpen
  \bibfield  {author} {\bibinfo {author} {\bibfnamefont {X.}~\bibnamefont
  {Meng}}, \bibinfo {author} {\bibfnamefont {A.}~\bibnamefont {Taylor-Noonan}},
  \bibinfo {author} {\bibfnamefont {C.}~\bibnamefont {Johnson}}, \bibinfo
  {author} {\bibfnamefont {W.~A.}\ \bibnamefont {Take}}, \bibinfo {author}
  {\bibfnamefont {E.}~\bibnamefont {Bowman}}, \ and\ \bibinfo {author}
  {\bibfnamefont {J.}~\bibnamefont {Gray}},\ }\bibfield  {title} {\enquote
  {\bibinfo {title} {Granular-fluid avalanches: the role of vertical structure
  and velocity shear},}\ }\href {\doibase
  https://doi.org/10.1017/jfm.2023.1022} {\bibfield  {journal} {\bibinfo
  {journal} {J. Fluid Mech.}\ }\textbf {\bibinfo {volume} {980}},\ \bibinfo
  {pages} {A11} (\bibinfo {year} {2024})}\BibitemShut {NoStop}%
\bibitem [{\citenamefont {Jerolmack}\ and\ \citenamefont
  {Daniels}(2019)}]{jerolmack2019viewing}%
  \BibitemOpen
  \bibfield  {author} {\bibinfo {author} {\bibfnamefont {D.~J.}\ \bibnamefont
  {Jerolmack}}\ and\ \bibinfo {author} {\bibfnamefont {K.~E.}\ \bibnamefont
  {Daniels}},\ }\bibfield  {title} {\enquote {\bibinfo {title} {Viewing
  earth’s surface as a soft-matter landscape},}\ }\href {\doibase
  https://doi.org/10.1038/s42254-019-0111-x} {\bibfield  {journal} {\bibinfo
  {journal} {Nat. Rev. Phys.}\ }\textbf {\bibinfo {volume} {1}},\ \bibinfo
  {pages} {716--730} (\bibinfo {year} {2019})}\BibitemShut {NoStop}%
\bibitem [{\citenamefont {Coussot}\ and\ \citenamefont
  {Ancey}(1999)}]{coussot1999rheophysical}%
  \BibitemOpen
  \bibfield  {author} {\bibinfo {author} {\bibfnamefont {P.}~\bibnamefont
  {Coussot}}\ and\ \bibinfo {author} {\bibfnamefont {C.}~\bibnamefont
  {Ancey}},\ }\bibfield  {title} {\enquote {\bibinfo {title} {Rheophysical
  classification of concentrated suspensions and granular pastes},}\ }\href
  {\doibase https://doi.org/10.1103/PhysRevE.59.4445} {\bibfield  {journal}
  {\bibinfo  {journal} {Phys. Rev. E}\ }\textbf {\bibinfo {volume} {59}},\
  \bibinfo {pages} {4445} (\bibinfo {year} {1999})}\BibitemShut {NoStop}%
\bibitem [{\citenamefont {Bancroft}\ and\ \citenamefont
  {Johnson}(2021)}]{bancroft2021drag}%
  \BibitemOpen
  \bibfield  {author} {\bibinfo {author} {\bibfnamefont {R.~S.}\ \bibnamefont
  {Bancroft}}\ and\ \bibinfo {author} {\bibfnamefont {C.~G.}\ \bibnamefont
  {Johnson}},\ }\bibfield  {title} {\enquote {\bibinfo {title} {Drag, diffusion
  and segregation in inertial granular flows},}\ }\href {\doibase
  https://doi.org/10.1017/jfm.2021.560} {\bibfield  {journal} {\bibinfo
  {journal} {J. Fluid Mech.}\ }\textbf {\bibinfo {volume} {924}},\ \bibinfo
  {pages} {A3} (\bibinfo {year} {2021})}\BibitemShut {NoStop}%
\bibitem [{\citenamefont {Kloss}\ \emph {et~al.}(2012)\citenamefont {Kloss},
  \citenamefont {Goniva}, \citenamefont {Hager}, \citenamefont {Amberger},\
  and\ \citenamefont {Pirker}}]{kloss2012models}%
  \BibitemOpen
  \bibfield  {author} {\bibinfo {author} {\bibfnamefont {C.}~\bibnamefont
  {Kloss}}, \bibinfo {author} {\bibfnamefont {C.}~\bibnamefont {Goniva}},
  \bibinfo {author} {\bibfnamefont {A.}~\bibnamefont {Hager}}, \bibinfo
  {author} {\bibfnamefont {S.}~\bibnamefont {Amberger}}, \ and\ \bibinfo
  {author} {\bibfnamefont {S.}~\bibnamefont {Pirker}},\ }\bibfield  {title}
  {\enquote {\bibinfo {title} {Models, algorithms and validation for opensource
  dem and cfd--dem},}\ }\href {\doibase
  https://doi.org/10.1504/PCFD.2012.047457} {\bibfield  {journal} {\bibinfo
  {journal} {Prog. Comput. Fluid Dyn.}\ }\textbf {\bibinfo {volume} {12}},\
  \bibinfo {pages} {140--152} (\bibinfo {year} {2012})}\BibitemShut {NoStop}%
\bibitem [{\citenamefont {Jiang}\ \emph {et~al.}(2018)\citenamefont {Jiang},
  \citenamefont {Fan}, \citenamefont {Li},\ and\ \citenamefont
  {Xiao}}]{jiang2018influence}%
  \BibitemOpen
  \bibfield  {author} {\bibinfo {author} {\bibfnamefont {Y.-J.}\ \bibnamefont
  {Jiang}}, \bibinfo {author} {\bibfnamefont {X.-Y.}\ \bibnamefont {Fan}},
  \bibinfo {author} {\bibfnamefont {T.-H.}\ \bibnamefont {Li}}, \ and\ \bibinfo
  {author} {\bibfnamefont {S.-Y.}\ \bibnamefont {Xiao}},\ }\bibfield  {title}
  {\enquote {\bibinfo {title} {Influence of particle-size segregation on the
  impact of dry granular flow},}\ }\href {\doibase
  https://doi.org/10.1016/j.powtec.2018.09.014} {\bibfield  {journal} {\bibinfo
   {journal} {Powder Technology}\ }\textbf {\bibinfo {volume} {340}},\ \bibinfo
  {pages} {39--51} (\bibinfo {year} {2018})}\BibitemShut {NoStop}%
\bibitem [{\citenamefont {Ferdowsi}\ \emph {et~al.}(2017)\citenamefont
  {Ferdowsi}, \citenamefont {Ortiz}, \citenamefont {Houssais},\ and\
  \citenamefont {Jerolmack}}]{ferdowsi2017river}%
  \BibitemOpen
  \bibfield  {author} {\bibinfo {author} {\bibfnamefont {B.}~\bibnamefont
  {Ferdowsi}}, \bibinfo {author} {\bibfnamefont {C.~P.}\ \bibnamefont {Ortiz}},
  \bibinfo {author} {\bibfnamefont {M.}~\bibnamefont {Houssais}}, \ and\
  \bibinfo {author} {\bibfnamefont {D.~J.}\ \bibnamefont {Jerolmack}},\
  }\bibfield  {title} {\enquote {\bibinfo {title} {River-bed armouring as a
  granular segregation phenomenon},}\ }\href {\doibase
  https://doi.org/10.1038/s41467-017-01681-3} {\bibfield  {journal} {\bibinfo
  {journal} {Nature communications}\ }\textbf {\bibinfo {volume} {8}},\
  \bibinfo {pages} {1363} (\bibinfo {year} {2017})}\BibitemShut {NoStop}%
\bibitem [{\citenamefont {Zhu}\ \emph {et~al.}(2007)\citenamefont {Zhu},
  \citenamefont {Zhou}, \citenamefont {Yang},\ and\ \citenamefont
  {Yu}}]{zhu2007discrete}%
  \BibitemOpen
  \bibfield  {author} {\bibinfo {author} {\bibfnamefont {H.~P.}\ \bibnamefont
  {Zhu}}, \bibinfo {author} {\bibfnamefont {Z.~Y.}\ \bibnamefont {Zhou}},
  \bibinfo {author} {\bibfnamefont {R.}~\bibnamefont {Yang}}, \ and\ \bibinfo
  {author} {\bibfnamefont {A.}~\bibnamefont {Yu}},\ }\bibfield  {title}
  {\enquote {\bibinfo {title} {Discrete particle simulation of particulate
  systems: theoretical developments},}\ }\href {\doibase
  https://doi.org/10.1016/j.ces.2006.12.089} {\bibfield  {journal} {\bibinfo
  {journal} {Chem. Eng. Sci.}\ }\textbf {\bibinfo {volume} {62}},\ \bibinfo
  {pages} {3378--3396} (\bibinfo {year} {2007})}\BibitemShut {NoStop}%
\bibitem [{\citenamefont {Trewhela}(2024)}]{trewhela2024segregation}%
  \BibitemOpen
  \bibfield  {author} {\bibinfo {author} {\bibfnamefont {T.}~\bibnamefont
  {Trewhela}},\ }\bibfield  {title} {\enquote {\bibinfo {title}
  {Segregation--rheology feedback in bidisperse granular flows: a coupled
  stokes’ problem},}\ }\href {\doibase https://doi.org/10.1017/jfm.2024.168}
  {\bibfield  {journal} {\bibinfo  {journal} {J. Fluid Mech.}\ }\textbf
  {\bibinfo {volume} {983}},\ \bibinfo {pages} {A45} (\bibinfo {year}
  {2024})}\BibitemShut {NoStop}%
\bibitem [{\citenamefont {Pusey}\ \emph {et~al.}()\citenamefont {Pusey},
  \citenamefont {Zaccarelli}, \citenamefont {Valeriani}, \citenamefont {Sanz},
  \citenamefont {Poon},\ and\ \citenamefont {Cates}}]{pusey2009hard}%
  \BibitemOpen
  \bibfield  {author} {\bibinfo {author} {\bibfnamefont {P.}~\bibnamefont
  {Pusey}}, \bibinfo {author} {\bibfnamefont {E.}~\bibnamefont {Zaccarelli}},
  \bibinfo {author} {\bibfnamefont {C.}~\bibnamefont {Valeriani}}, \bibinfo
  {author} {\bibfnamefont {E.}~\bibnamefont {Sanz}}, \bibinfo {author}
  {\bibfnamefont {W.~C.}\ \bibnamefont {Poon}}, \ and\ \bibinfo {author}
  {\bibfnamefont {M.~E.}\ \bibnamefont {Cates}},\ }\bibfield  {title} {\enquote
  {\bibinfo {title} {Hard spheres: crystallization and glass formation},}\
  }\href {\doibase https://doi.org/10.1098/rsta.2009.0181} {\
  https://doi.org/10.1098/rsta.2009.0181}\BibitemShut {NoStop}%
\bibitem [{\citenamefont {Dziugys}\ and\ \citenamefont
  {Navakas}(2007)}]{dziugys2007influence}%
  \BibitemOpen
  \bibfield  {author} {\bibinfo {author} {\bibfnamefont {A.}~\bibnamefont
  {Dziugys}}\ and\ \bibinfo {author} {\bibfnamefont {R.}~\bibnamefont
  {Navakas}},\ }\bibfield  {title} {\enquote {\bibinfo {title} {Influence of
  mechanical properties on mixing and segregation of granular material},}\
  }\href {\doibase 10.1016/j.powtec.2007.04.019} {\bibfield  {journal}
  {\bibinfo  {journal} {Powder Technology}\ }\textbf {\bibinfo {volume}
  {189}},\ \bibinfo {pages} {485--491} (\bibinfo {year} {2007})}\BibitemShut
  {NoStop}%
\bibitem [{\citenamefont {Li}\ \emph {et~al.}(2022)\citenamefont {Li},
  \citenamefont {Dong}, \citenamefont {Liu},\ and\ \citenamefont
  {Shen}}]{li2022dem}%
  \BibitemOpen
  \bibfield  {author} {\bibinfo {author} {\bibfnamefont {C.}~\bibnamefont
  {Li}}, \bibinfo {author} {\bibfnamefont {K.}~\bibnamefont {Dong}}, \bibinfo
  {author} {\bibfnamefont {S.}~\bibnamefont {Liu}}, \ and\ \bibinfo {author}
  {\bibfnamefont {Y.}~\bibnamefont {Shen}},\ }\bibfield  {title} {\enquote
  {\bibinfo {title} {Dem study of particle segregation in the throat region of
  a blast furnace},}\ }\href {\doibase 10.1016/j.powtec.2022.117310} {\bibfield
   {journal} {\bibinfo  {journal} {Powder Technology}\ }\textbf {\bibinfo
  {volume} {399}},\ \bibinfo {pages} {117310} (\bibinfo {year}
  {2022})}\BibitemShut {NoStop}%
\bibitem [{\citenamefont {Mueth}(2003)}]{mueth2003measurements}%
  \BibitemOpen
  \bibfield  {author} {\bibinfo {author} {\bibfnamefont {D.~M.}\ \bibnamefont
  {Mueth}},\ }\bibfield  {title} {\enquote {\bibinfo {title} {Measurements of
  particle dynamics in slow, dense granular couette flow},}\ }\href {\doibase
  https://doi.org/10.1103/PhysRevE.67.011304} {\bibfield  {journal} {\bibinfo
  {journal} {Phys. Rev. E}\ }\textbf {\bibinfo {volume} {67}},\ \bibinfo
  {pages} {011304} (\bibinfo {year} {2003})}\BibitemShut {NoStop}%
\bibitem [{\citenamefont {May}\ \emph {et~al.}(2010)\citenamefont {May},
  \citenamefont {Golick}, \citenamefont {Phillips}, \citenamefont {Shearer},\
  and\ \citenamefont {Daniels}}]{may2010shear}%
  \BibitemOpen
  \bibfield  {author} {\bibinfo {author} {\bibfnamefont {L.~B.}\ \bibnamefont
  {May}}, \bibinfo {author} {\bibfnamefont {L.~A.}\ \bibnamefont {Golick}},
  \bibinfo {author} {\bibfnamefont {K.~C.}\ \bibnamefont {Phillips}}, \bibinfo
  {author} {\bibfnamefont {M.}~\bibnamefont {Shearer}}, \ and\ \bibinfo
  {author} {\bibfnamefont {K.~E.}\ \bibnamefont {Daniels}},\ }\bibfield
  {title} {\enquote {\bibinfo {title} {Shear-driven size segregation of
  granular materials: modeling and experiment},}\ }\href {\doibase
  https://doi.org/10.1103/PhysRevE.81.051301} {\bibfield  {journal} {\bibinfo
  {journal} {Phys. Rev. E}\ }\textbf {\bibinfo {volume} {81}},\ \bibinfo
  {pages} {051301} (\bibinfo {year} {2010})}\BibitemShut {NoStop}%
\bibitem [{\citenamefont {Artoni}\ \emph {et~al.}(2018)\citenamefont {Artoni},
  \citenamefont {Soligo}, \citenamefont {Paul},\ and\ \citenamefont
  {Richard}}]{artoni2018shear}%
  \BibitemOpen
  \bibfield  {author} {\bibinfo {author} {\bibfnamefont {R.}~\bibnamefont
  {Artoni}}, \bibinfo {author} {\bibfnamefont {A.}~\bibnamefont {Soligo}},
  \bibinfo {author} {\bibfnamefont {J.-M.}\ \bibnamefont {Paul}}, \ and\
  \bibinfo {author} {\bibfnamefont {P.}~\bibnamefont {Richard}},\ }\bibfield
  {title} {\enquote {\bibinfo {title} {Shear localization and wall friction in
  confined dense granular flows},}\ }\href {\doibase
  https://doi.org/10.1017/jfm.2018.407} {\bibfield  {journal} {\bibinfo
  {journal} {J. Fluid Mech.}\ }\textbf {\bibinfo {volume} {849}},\ \bibinfo
  {pages} {395--418} (\bibinfo {year} {2018})}\BibitemShut {NoStop}%
\bibitem [{\citenamefont {Rognon}\ \emph {et~al.}(2007)\citenamefont {Rognon},
  \citenamefont {Roux}, \citenamefont {Naa{\"\i}m},\ and\ \citenamefont
  {Chevoir}}]{rognon2007dense}%
  \BibitemOpen
  \bibfield  {author} {\bibinfo {author} {\bibfnamefont {P.~G.}\ \bibnamefont
  {Rognon}}, \bibinfo {author} {\bibfnamefont {J.-N.}\ \bibnamefont {Roux}},
  \bibinfo {author} {\bibfnamefont {M.}~\bibnamefont {Naa{\"\i}m}}, \ and\
  \bibinfo {author} {\bibfnamefont {F.}~\bibnamefont {Chevoir}},\ }\bibfield
  {title} {\enquote {\bibinfo {title} {Dense flows of bidisperse assemblies of
  disks down an inclined plane},}\ }\href {\doibase
  https://doi.org/10.1063/1.2722242} {\bibfield  {journal} {\bibinfo  {journal}
  {Phys. Fluids}\ }\textbf {\bibinfo {volume} {19}} (\bibinfo {year} {2007}),\
  https://doi.org/10.1063/1.2722242}\BibitemShut {NoStop}%
\bibitem [{\citenamefont {Trewhela}\ and\ \citenamefont
  {Ancey}(2021)}]{trewhela2021conveyor}%
  \BibitemOpen
  \bibfield  {author} {\bibinfo {author} {\bibfnamefont {T.}~\bibnamefont
  {Trewhela}}\ and\ \bibinfo {author} {\bibfnamefont {C.}~\bibnamefont
  {Ancey}},\ }\bibfield  {title} {\enquote {\bibinfo {title} {A conveyor belt
  experimental setup to study the internal dynamics of granular avalanches},}\
  }\href {\doibase https://doi.org/10.1007/s00348-021-03299-0} {\bibfield
  {journal} {\bibinfo  {journal} {Exp. Fluids}\ }\textbf {\bibinfo {volume}
  {62}},\ \bibinfo {pages} {1--17} (\bibinfo {year} {2021})}\BibitemShut
  {NoStop}%
\bibitem [{\citenamefont {Pouliquen}\ \emph {et~al.}(2006)\citenamefont
  {Pouliquen}, \citenamefont {Cassar}, \citenamefont {Jop}, \citenamefont
  {Forterre},\ and\ \citenamefont {Nicolas}}]{pouliquen2006flow}%
  \BibitemOpen
  \bibfield  {author} {\bibinfo {author} {\bibfnamefont {O.}~\bibnamefont
  {Pouliquen}}, \bibinfo {author} {\bibfnamefont {C.}~\bibnamefont {Cassar}},
  \bibinfo {author} {\bibfnamefont {P.}~\bibnamefont {Jop}}, \bibinfo {author}
  {\bibfnamefont {Y.}~\bibnamefont {Forterre}}, \ and\ \bibinfo {author}
  {\bibfnamefont {M.}~\bibnamefont {Nicolas}},\ }\bibfield  {title} {\enquote
  {\bibinfo {title} {Flow of dense granular material: towards simple
  constitutive laws},}\ }\href {\doibase
  https://doi.org/10.1088/1742-5468/2006/07/P07020} {\bibfield  {journal}
  {\bibinfo  {journal} {J. Stat. Mech.: Theory Exp.}\ }\textbf {\bibinfo
  {volume} {2006}},\ \bibinfo {pages} {P07020} (\bibinfo {year}
  {2006})}\BibitemShut {NoStop}%
\bibitem [{\citenamefont {van~der Vaart}\ \emph {et~al.}(2018)\citenamefont
  {van~der Vaart}, \citenamefont {van Schrojenstein~Lantman}, \citenamefont
  {Weinhart}, \citenamefont {Luding}, \citenamefont {Ancey},\ and\
  \citenamefont {Thornton}}]{van2018segregation}%
  \BibitemOpen
  \bibfield  {author} {\bibinfo {author} {\bibfnamefont {K.}~\bibnamefont
  {van~der Vaart}}, \bibinfo {author} {\bibfnamefont {M.}~\bibnamefont {van
  Schrojenstein~Lantman}}, \bibinfo {author} {\bibfnamefont {T.}~\bibnamefont
  {Weinhart}}, \bibinfo {author} {\bibfnamefont {S.}~\bibnamefont {Luding}},
  \bibinfo {author} {\bibfnamefont {C.}~\bibnamefont {Ancey}}, \ and\ \bibinfo
  {author} {\bibfnamefont {A.}~\bibnamefont {Thornton}},\ }\bibfield  {title}
  {\enquote {\bibinfo {title} {Segregation of large particles in dense granular
  flows suggests a granular saffman effect},}\ }\href {\doibase
  https://doi.org/10.1103/PhysRevFluids.3.074303} {\bibfield  {journal}
  {\bibinfo  {journal} {Phys. Rev. Fluids}\ }\textbf {\bibinfo {volume} {3}},\
  \bibinfo {pages} {074303} (\bibinfo {year} {2018})}\BibitemShut {NoStop}%
\bibitem [{\citenamefont {Jop}, \citenamefont {Forterre},\ and\ \citenamefont
  {Pouliquen}(2006)}]{jop2006constitutive}%
  \BibitemOpen
  \bibfield  {author} {\bibinfo {author} {\bibfnamefont {P.}~\bibnamefont
  {Jop}}, \bibinfo {author} {\bibfnamefont {Y.}~\bibnamefont {Forterre}}, \
  and\ \bibinfo {author} {\bibfnamefont {O.}~\bibnamefont {Pouliquen}},\
  }\bibfield  {title} {\enquote {\bibinfo {title} {A constitutive law for dense
  granular flows},}\ }\href {\doibase https://doi.org/10.1038/nature04801}
  {\bibfield  {journal} {\bibinfo  {journal} {Nature}\ }\textbf {\bibinfo
  {volume} {441}},\ \bibinfo {pages} {727--730} (\bibinfo {year}
  {2006})}\BibitemShut {NoStop}%
\bibitem [{\citenamefont {Utter}\ and\ \citenamefont
  {Behringer}(2004)}]{utter2004self}%
  \BibitemOpen
  \bibfield  {author} {\bibinfo {author} {\bibfnamefont {B.}~\bibnamefont
  {Utter}}\ and\ \bibinfo {author} {\bibfnamefont {R.~P.}\ \bibnamefont
  {Behringer}},\ }\bibfield  {title} {\enquote {\bibinfo {title}
  {Self-diffusion in dense granular shear flows},}\ }\href {\doibase
  https://doi.org/10.1103/PhysRevE.69.031308} {\bibfield  {journal} {\bibinfo
  {journal} {Phys. Rev. E}\ }\textbf {\bibinfo {volume} {69}},\ \bibinfo
  {pages} {031308} (\bibinfo {year} {2004})}\BibitemShut {NoStop}%
\bibitem [{\citenamefont {Maguire}\ \emph {et~al.}(2024)\citenamefont
  {Maguire}, \citenamefont {Barker}, \citenamefont {Rauter}, \citenamefont
  {Johnson},\ and\ \citenamefont {Gray}}]{maguire2024particle}%
  \BibitemOpen
  \bibfield  {author} {\bibinfo {author} {\bibfnamefont {E.}~\bibnamefont
  {Maguire}}, \bibinfo {author} {\bibfnamefont {T.}~\bibnamefont {Barker}},
  \bibinfo {author} {\bibfnamefont {M.}~\bibnamefont {Rauter}}, \bibinfo
  {author} {\bibfnamefont {C.}~\bibnamefont {Johnson}}, \ and\ \bibinfo
  {author} {\bibfnamefont {J.}~\bibnamefont {Gray}},\ }\bibfield  {title}
  {\enquote {\bibinfo {title} {Particle-size segregation patterns in a
  partially filled triangular rotating drum},}\ }\href {\doibase
  https://doi.org/10.1017/jfm.2023.1022} {\bibfield  {journal} {\bibinfo
  {journal} {J. Fluid Mech.}\ }\textbf {\bibinfo {volume} {979}},\ \bibinfo
  {pages} {A40} (\bibinfo {year} {2024})}\BibitemShut {NoStop}%
\bibitem [{\citenamefont {Kumawat}, \citenamefont {Sahu},\ and\ \citenamefont
  {Tripathi}(2025)}]{kumawat2024transient}%
  \BibitemOpen
  \bibfield  {author} {\bibinfo {author} {\bibfnamefont {S.}~\bibnamefont
  {Kumawat}}, \bibinfo {author} {\bibfnamefont {V.~K.}\ \bibnamefont {Sahu}}, \
  and\ \bibinfo {author} {\bibfnamefont {A.}~\bibnamefont {Tripathi}},\
  }\bibfield  {title} {\enquote {\bibinfo {title} {Transient segregation of
  different density granular mixtures},}\ }\href {\doibase
  https://doi.org/10.1017/jfm.2025.207} {\bibfield  {journal} {\bibinfo
  {journal} {J. Fluid Mech.}\ }\textbf {\bibinfo {volume} {1008}},\ \bibinfo
  {pages} {A53} (\bibinfo {year} {2025})}\BibitemShut {NoStop}%
\bibitem [{\citenamefont {Zhao}\ \emph {et~al.}()\citenamefont {Zhao},
  \citenamefont {Noto}, \citenamefont {Li}, \citenamefont {Trewhela},\ and\
  \citenamefont {Ulloa}}]{Zhao_zenodo_2025}%
  \BibitemOpen
  \bibfield  {author} {\bibinfo {author} {\bibfnamefont {T.}~\bibnamefont
  {Zhao}}, \bibinfo {author} {\bibfnamefont {D.}~\bibnamefont {Noto}}, \bibinfo
  {author} {\bibfnamefont {X.}~\bibnamefont {Li}}, \bibinfo {author}
  {\bibfnamefont {T.}~\bibnamefont {Trewhela}}, \ and\ \bibinfo {author}
  {\bibfnamefont {H.}~\bibnamefont {Ulloa}},\ }\bibfield  {title} {\enquote
  {\bibinfo {title} {Dataset: Scaling particle-size segregation in wide-ranging
  sheared granular flows},}\ }\href {\doibase
  https://doi.org/10.5281/zenodo.15272629} {\bibfield  {journal} {\bibinfo
  {journal} {Zenodo}\ }https://doi.org/10.5281/zenodo.15272629}\BibitemShut
  {NoStop}%
\bibitem [{\citenamefont {Thornton}\ \emph {et~al.}(2012)\citenamefont
  {Thornton}, \citenamefont {Weinhart}, \citenamefont {Luding},\ and\
  \citenamefont {Bokhove}}]{thornton2012modeling}%
  \BibitemOpen
  \bibfield  {author} {\bibinfo {author} {\bibfnamefont {A.}~\bibnamefont
  {Thornton}}, \bibinfo {author} {\bibfnamefont {T.}~\bibnamefont {Weinhart}},
  \bibinfo {author} {\bibfnamefont {S.}~\bibnamefont {Luding}}, \ and\ \bibinfo
  {author} {\bibfnamefont {O.}~\bibnamefont {Bokhove}},\ }\bibfield  {title}
  {\enquote {\bibinfo {title} {Modeling of particle size segregation:
  calibration using the discrete particle method},}\ }\href {\doibase
  https://doi.org/10.1142/S0129183112400141} {\bibfield  {journal} {\bibinfo
  {journal} {International journal of modern physics C}\ }\textbf {\bibinfo
  {volume} {23}},\ \bibinfo {pages} {1240014} (\bibinfo {year}
  {2012})}\BibitemShut {NoStop}%
\bibitem [{\citenamefont {Wiederseiner}\ \emph {et~al.}(2011)\citenamefont
  {Wiederseiner}, \citenamefont {Andreini}, \citenamefont {{\'E}pely-Chauvin},
  \citenamefont {Moser}, \citenamefont {Monnereau}, \citenamefont {Gray},\ and\
  \citenamefont {Ancey}}]{wiederseiner2011experimental}%
  \BibitemOpen
  \bibfield  {author} {\bibinfo {author} {\bibfnamefont {S.}~\bibnamefont
  {Wiederseiner}}, \bibinfo {author} {\bibfnamefont {N.}~\bibnamefont
  {Andreini}}, \bibinfo {author} {\bibfnamefont {G.}~\bibnamefont
  {{\'E}pely-Chauvin}}, \bibinfo {author} {\bibfnamefont {G.}~\bibnamefont
  {Moser}}, \bibinfo {author} {\bibfnamefont {M.}~\bibnamefont {Monnereau}},
  \bibinfo {author} {\bibfnamefont {J.}~\bibnamefont {Gray}}, \ and\ \bibinfo
  {author} {\bibfnamefont {C.}~\bibnamefont {Ancey}},\ }\bibfield  {title}
  {\enquote {\bibinfo {title} {Experimental investigation into segregating
  granular flows down chutes},}\ }\href {\doibase
  https://doi.org/10.1063/1.3536658} {\bibfield  {journal} {\bibinfo  {journal}
  {Physics of Fluids}\ }\textbf {\bibinfo {volume} {23}} (\bibinfo {year}
  {2011}),\ https://doi.org/10.1063/1.3536658}\BibitemShut {NoStop}%
\end{thebibliography}%

\end{document}